\documentclass[letterpaper]{emulateapj}

\usepackage{amsmath}
\usepackage{amssymb}
\usepackage{xspace}
\usepackage{color}
\usepackage{ifthen}
\usepackage{xspace}
\usepackage{graphicx}

\newcommand{\Msun}{{\ensuremath{\mathrm{M}_{\odot}}}\xspace}

\slugcomment{Draft for ApJ, \today}

\usepackage{ulem}
\begin{document}

\title{Dependence of \textsl{S}-Process Nucleosynthesis in Massive
  Stars on Triple-Alpha and $^{12}$C($\alpha,\gamma$)$^{16}$O Reaction Rate Uncertainties}

\author{Clarisse~Tur}
\affil{National Superconducting Cyclotron Laboratory,\\
Michigan State University, 1 Cyclotron Laboratory, East Lansing, MI 48824-1321;\\
Joint Institute for Nuclear Astrophysics} \email{tur@nscl.msu.edu}

\author{Alexander~Heger}
\affil{School of Physics and Astronomy, 
University of Minnesota, Twin Cities, 
Minneapolis, MN 55455-0149; \\
Theoretical Astrophysics Group, MS B227, Los Alamos
National Laboratory, Los Alamos, NM 87545;\\
Department of Astronomy and Astrophysics, University of
California, Santa Cruz, CA 95064; \\
Joint Institute for Nuclear Astrophysics} 
\email{alex@physics.umn.edu}

\author{Sam~M.~Austin}
\affil{National Superconducting Cyclotron Laboratory,\\
Michigan State University, 1 Cyclotron Laboratory, East Lansing, MI 48824-1321;\\
Joint Institute for Nuclear Astrophysics\phantom{.}}
\email{austin@nscl.msu.edu}

\begin{abstract}

We have studied the sensitivity of \textsl{s}-process nucleosynthesis
in massive stars to $\pm {2\sigma}$ variations in the rates of the
triple-$\alpha$ and $^{12}$C($\alpha,\gamma$)$^{16}$O reactions.  We
simulated the evolution of massive stars from H-burning through
Fe-core collapse, followed by a supernova explosion. We found that:
the production factors of \textsl{s}-process nuclides between
$^{58}$Fe and $^{96}$Zr change strongly with changes in the He burning
reaction rates; using the \citet{lod03} solar abundances rather than
those of \citet{and89} reduces \textsl{s}-process nucleosynthesis;
later burning phases beyond core He burning and shell C burning
have a significant effect on post-explosive production factors.  We
also discuss the implications of the uncertainties in the helium
burning rates for evidence of a new primary neutron capture process
(LEPP) in massive stars.

\end{abstract}


\section{Introduction}

Half of the elements between Fe and Bi are produced by the slow
(\textsl{s}) neutron capture process and most of the remainder by the
rapid (\textsl{r}) neutron capture process.  About 35 additional
neutron deficient stable nuclides above $^{56}$Fe, the
\textsl{p}-process nuclei, are produced in explosive
processes (\citealt{pra90a}).

Two components, the main and the weak \textsl{s}-processes, are required to
explain the  isotopic distributions of \textsl{s}-process nuclides. The main
\textsl{s}-process occurs in low-mass ($\lesssim4\,\Msun$) asymptotic giant
branch (AGB) stars and contributes mainly to the production of
heavier elements, with  a smaller contribution to $A \leq 90$. The
weak \textsl{s}-process occurs during the late evolutionary stages of massive 
stars ($\gtrsim 10\,\Msun$) and produces nuclides up to $A \simeq
90$.

Recently, \cite{tra04} summed the contributions of the
\textsl{r}-process, the main and weak \textsl{s}-process, and the
\textsl{p}-process to the abundances of Sr, Y, and Zr.  The summed
contributions were smaller than the observed solar abundances by 8\%,
18\%, and 18\%, respectively. They concluded that an additional light
element primary \textsl{s}-process contribution from massive stars
(LEPP) is needed to explain this difference; the nature and site of
the LEPP are unknown.  This LEPP has also been invoked by \cite{mon07}
to explain the abundances of a larger group of light
\textsl{r}-process elements.  Since the LEPP effects are relatively
small, and since some of the LEPP elements are produced with
relatively large abundance in the weak \textsl{s}-process, it is
important to establish whether the uncertainties in the weak
\textsl{s}-process are sufficiently small that the claim of LEPP
contributions is robust.

Nuclide production in the weak \textsl{s}-process also depends on the
rate of the neutron source $^{22}$Ne($\alpha,n$)$^{25}$Mg and on the
capture cross sections for the neutron poisons (medium-weight isotopes
up to Fe, including $^{12}$C, $^{16}$O, $^{20}$Ne, $^{22}$Ne,
$^{23}$Na, $^{24}$Mg, and $^{25}$Mg). Neither the source strength nor
the neutron capture cross sections for the poisons are known with
sufficient accuracy (\citealt{hei08}). We shall not deal with these
issues in this paper, but rather with the more indirect effects of
uncertainties in the rates ($R_{3\alpha}$ and $R_{\alpha,12}$) of the
triple alpha and $^{12}$C($\alpha,\gamma$)$^{16}$O reactions, and in
the initial stellar composition. For example, we have shown in a
previous paper (\citealt{tur07}; see also \citealt{wea93};
\citealt{woo03}; \citealt{woo07}), that the amount of the above
neutron poisons present during the weak \textsl{s}-process in massive
stars depends sensitively on these rates and the initial stellar composition.

Most earlier studies of the weak \textsl{s}-process focused on
production toward the end of core He burning by neutrons from the
$^{22}$Ne($\alpha,n$)$^{25}$Mg reaction (\citealt{cou74};
\citealt{lam77}; \citealt{arn85}; \citealt{bus85}; \citealt{lan89};
\citealt{pra90b}; \citealt{the00}; \citealt{rai91a};
\citealt{bar92}). Later papers considered also a second exposure at
higher temperatures and neutron densities peaking during shell C
burning (\citealt{rai91b}; \citealt{rai92}; \citealt{rai93};
\citealt{the07}).  Explosive processing in the supernova explosion
was not considered. Recent calculations of \textsl{s}-process yields have been
extended to consider the entire evolutionary history of the star,
including  explosive burning (\citealt{hof01}; \citealt{rau02};
\citealt{lim03}).

In this paper we simulate the evolution of massive stars from
H-burning through Fe-core collapse, followed by a supernova
explosion. Our principal purposes are (1) to establish the magnitude
of weak \textsl{s}-process production in a self-consistent model; (2)
to study the uncertainties in weak \textsl{s}-process nucleosynthesis
arising from uncertainties in $R_{3\alpha}$ and $R_{\alpha,12}$; (3)
to study the effects of different stellar abundances, specifically
those of \cite{lod03} and \cite{and89}, hereafter L03 and AG89; (4) to
delineate the stages of stellar evolution during which weak
\textsl{s}-process elements are produced; and (5) to assess the
bearing of these uncertainties on the robustness of the LEPP process.

In Section~2, we describe the stellar model and the input physics relevant 
to the treatment of the weak \textsl{s}-process. Section~3 gives our results for 
the dependence of the post-explosive weak-s process production factors on 
changes in the rates of the helium burning reactions and on different 
initial stellar abundances. In Section~4, we 
show the contribution to the weak \textsl{s}-process abundances of the various 
stellar burning stages prior to the supernova explosion.  In Section~5, we 
investigate the range of weak \textsl{s}-process production of Sr, Y, and Zr 
allowed by the uncertainties in the helium burning reactions.

\begin{figure*}
\centering
\includegraphics[angle=90,width=0.475\textwidth]{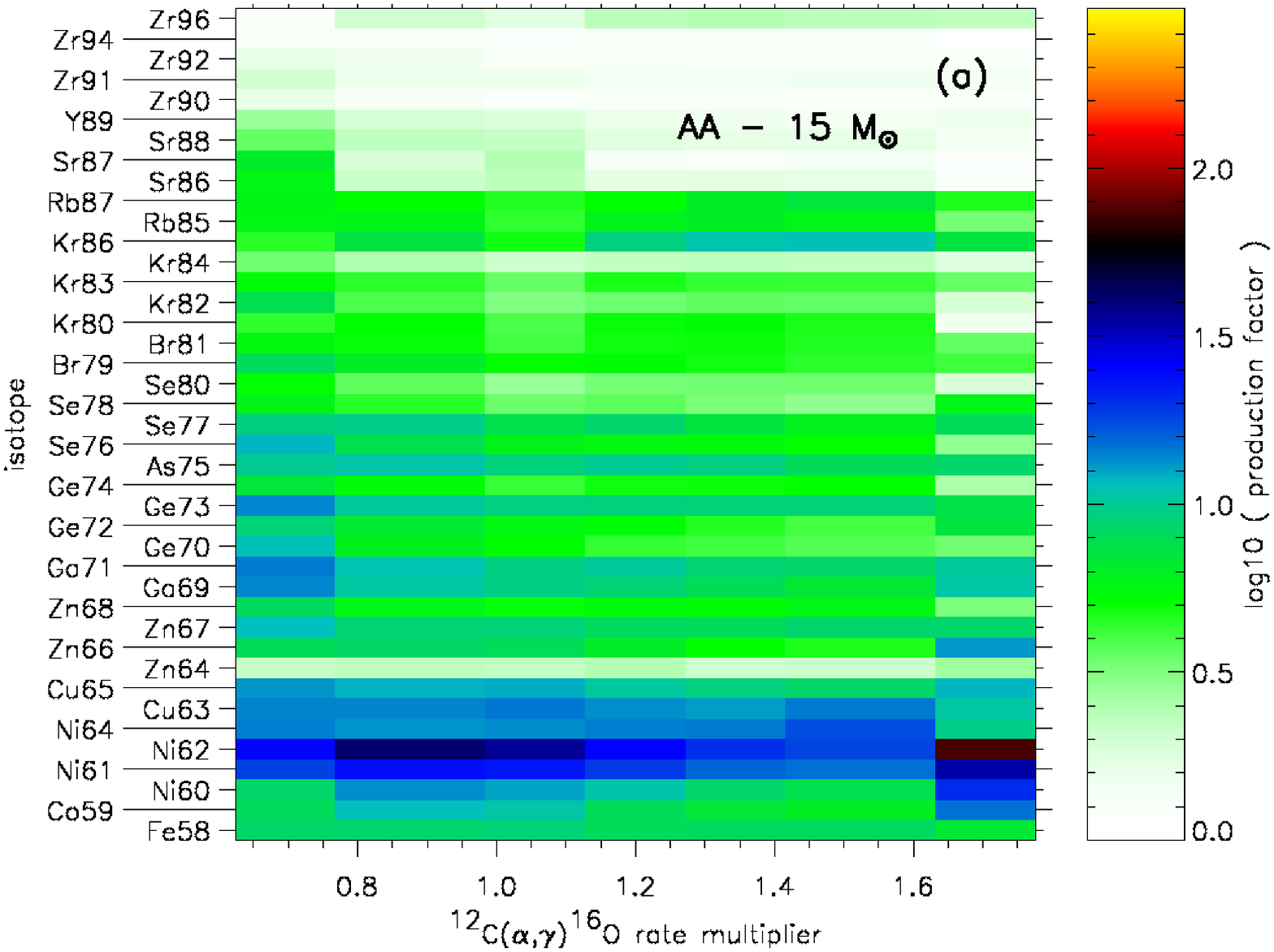}
\hfill
\includegraphics[angle=90,width=0.475\textwidth]{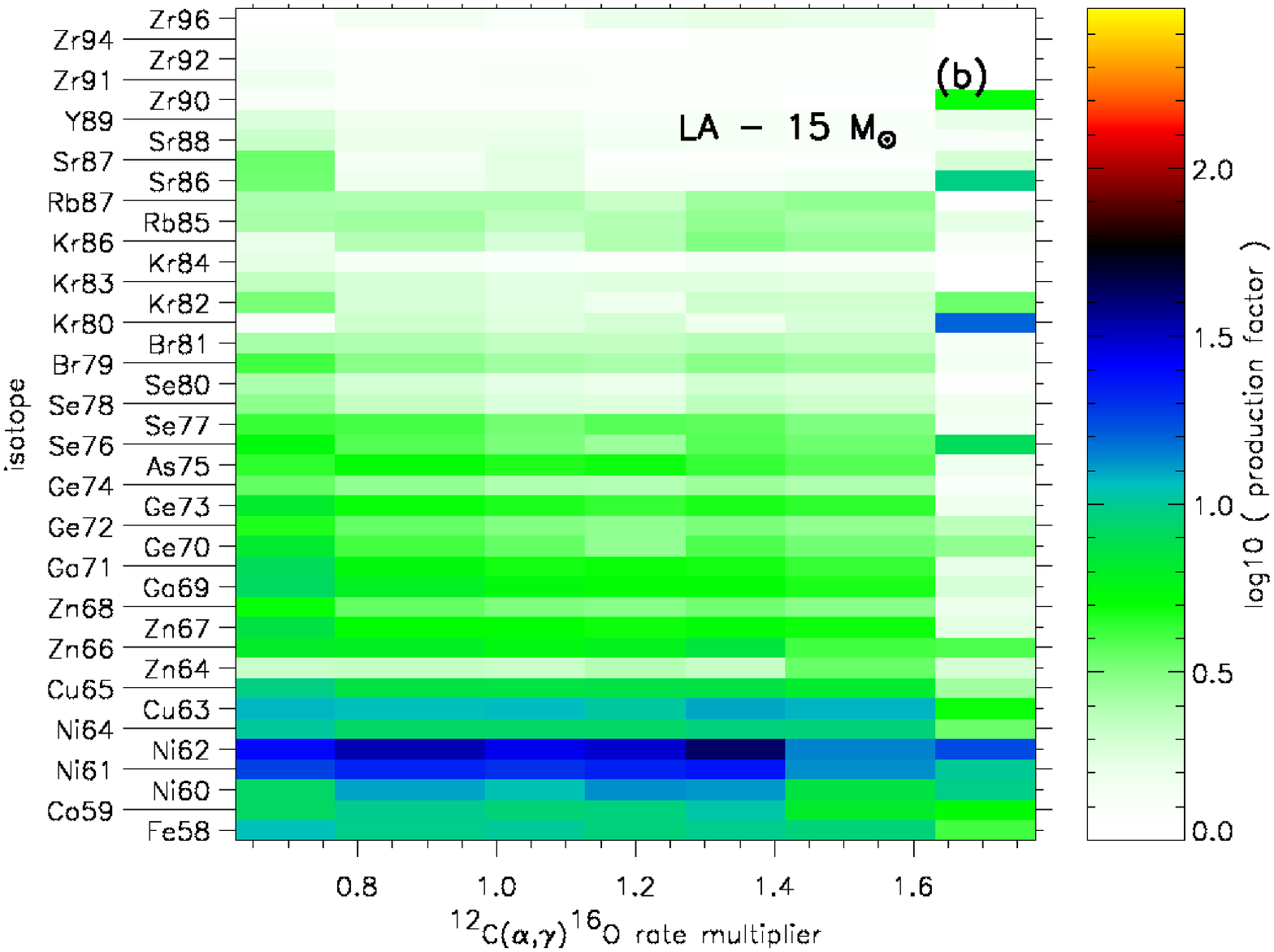}
\hfill
\includegraphics[angle=90,width=0.475\textwidth]{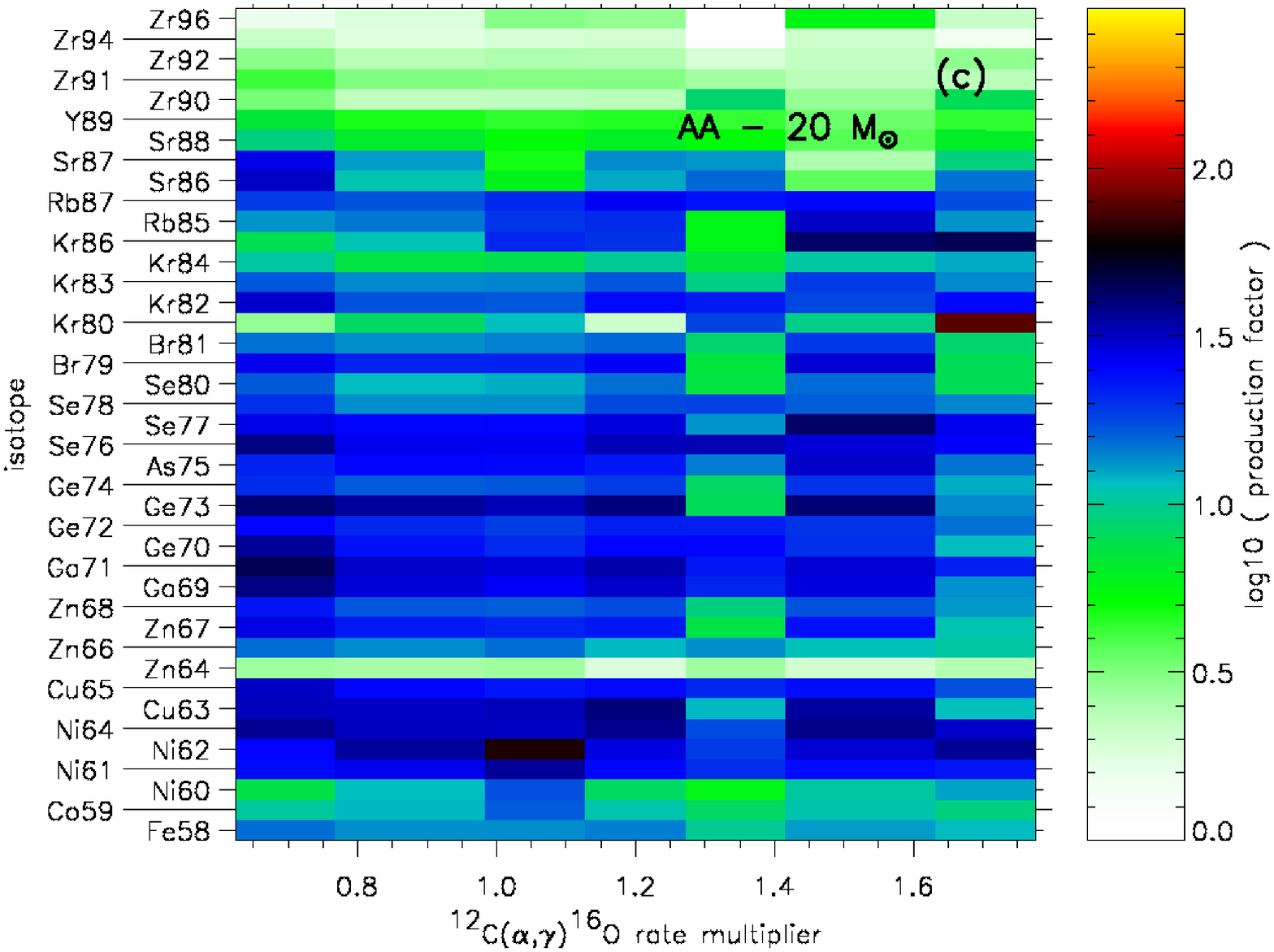}
\hfill
\includegraphics[angle=90,width=0.475\textwidth]{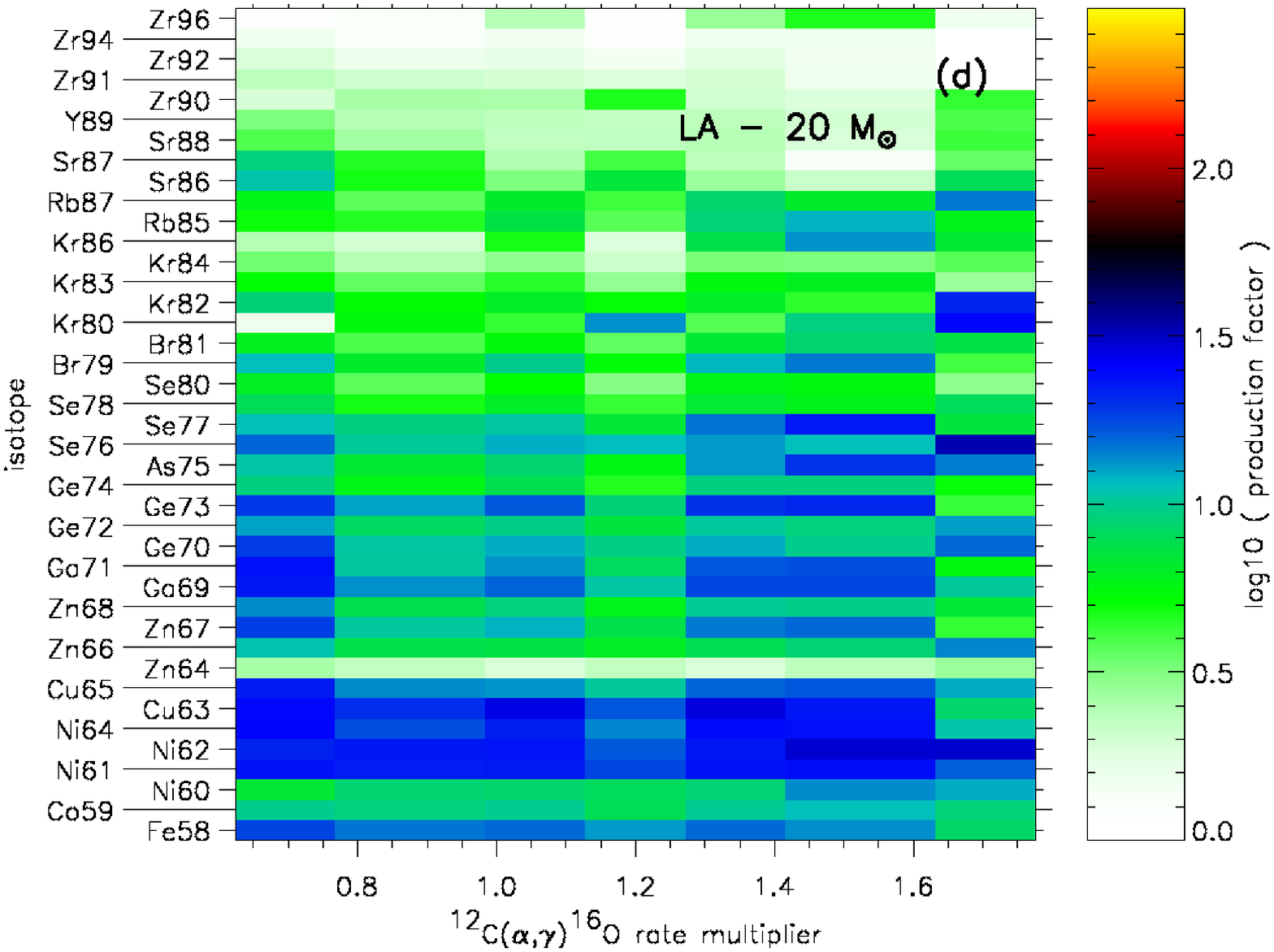}
\hfill
\includegraphics[angle=90,width=0.475\textwidth]{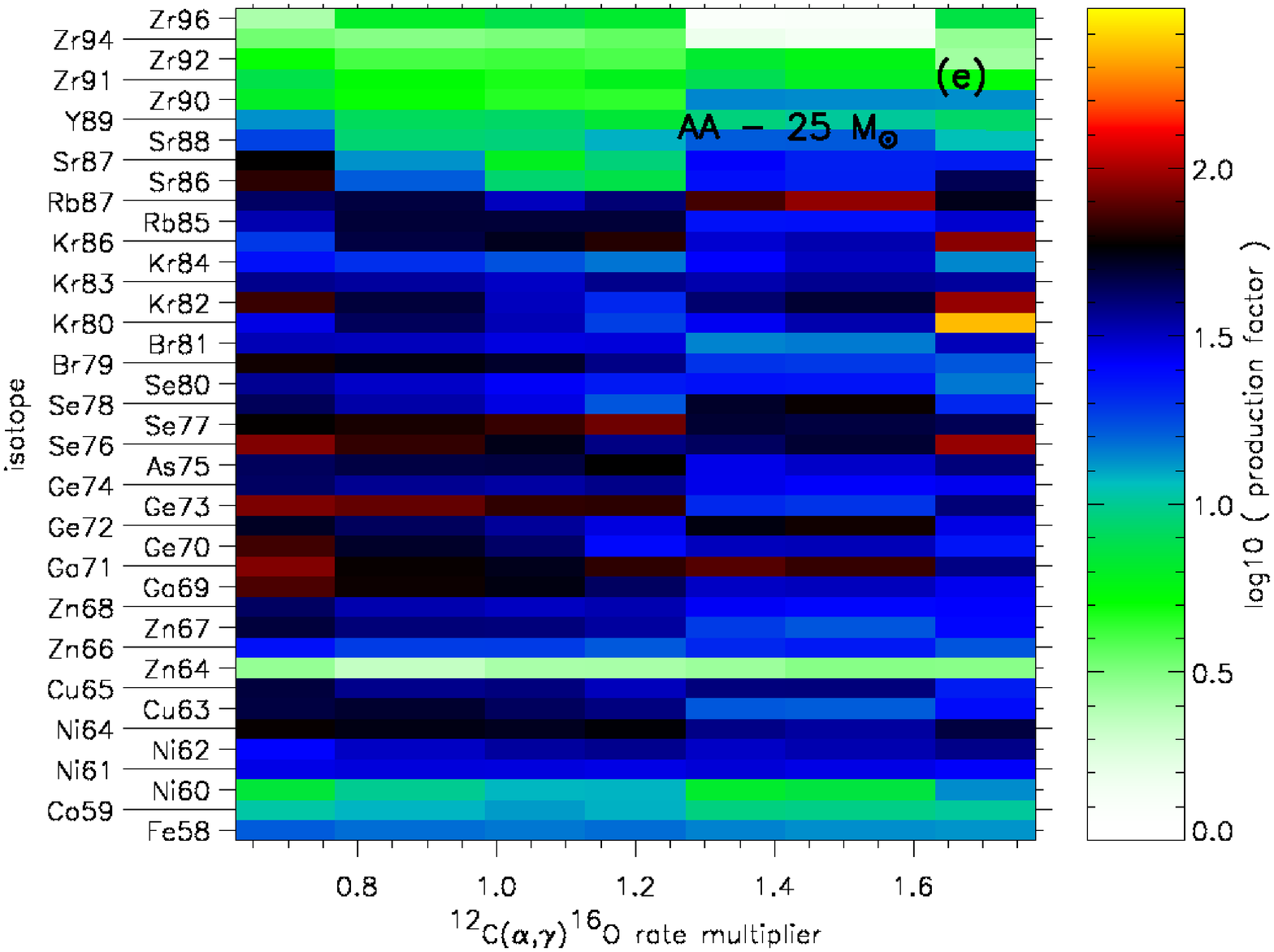}
\hfill
\includegraphics[angle=90,width=0.475\textwidth]{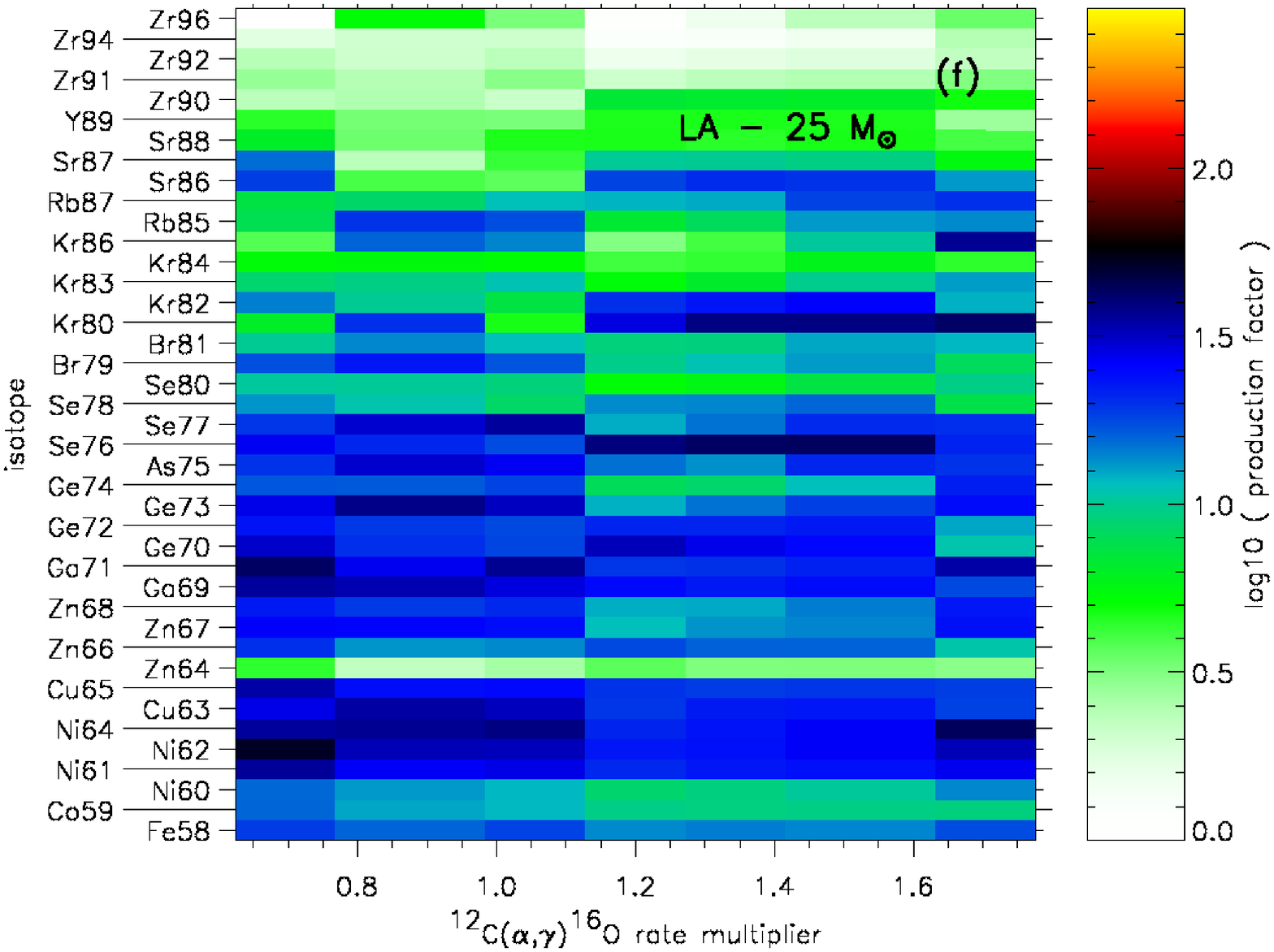}
\caption{Post-explosive production factors as a function of
  $R_{\alpha,12}$. \textbf{(a)} The AA series,
  $15\,\Msun$. \textbf{(b)} The LA series, $15\,\Msun$. \textbf{(c)}
  The AA series, $20\,\Msun$.  \textbf{(d)} The LA series,
  $20\,\Msun$. \textbf{(e)} The AA series, $25\,\Msun$. \textbf{(f)}
  The LA series, $25\,\Msun$.} \label{pfVs12cag}
\end{figure*}

\begin{figure*}
\centering
\includegraphics[angle=90,width=0.475\textwidth]{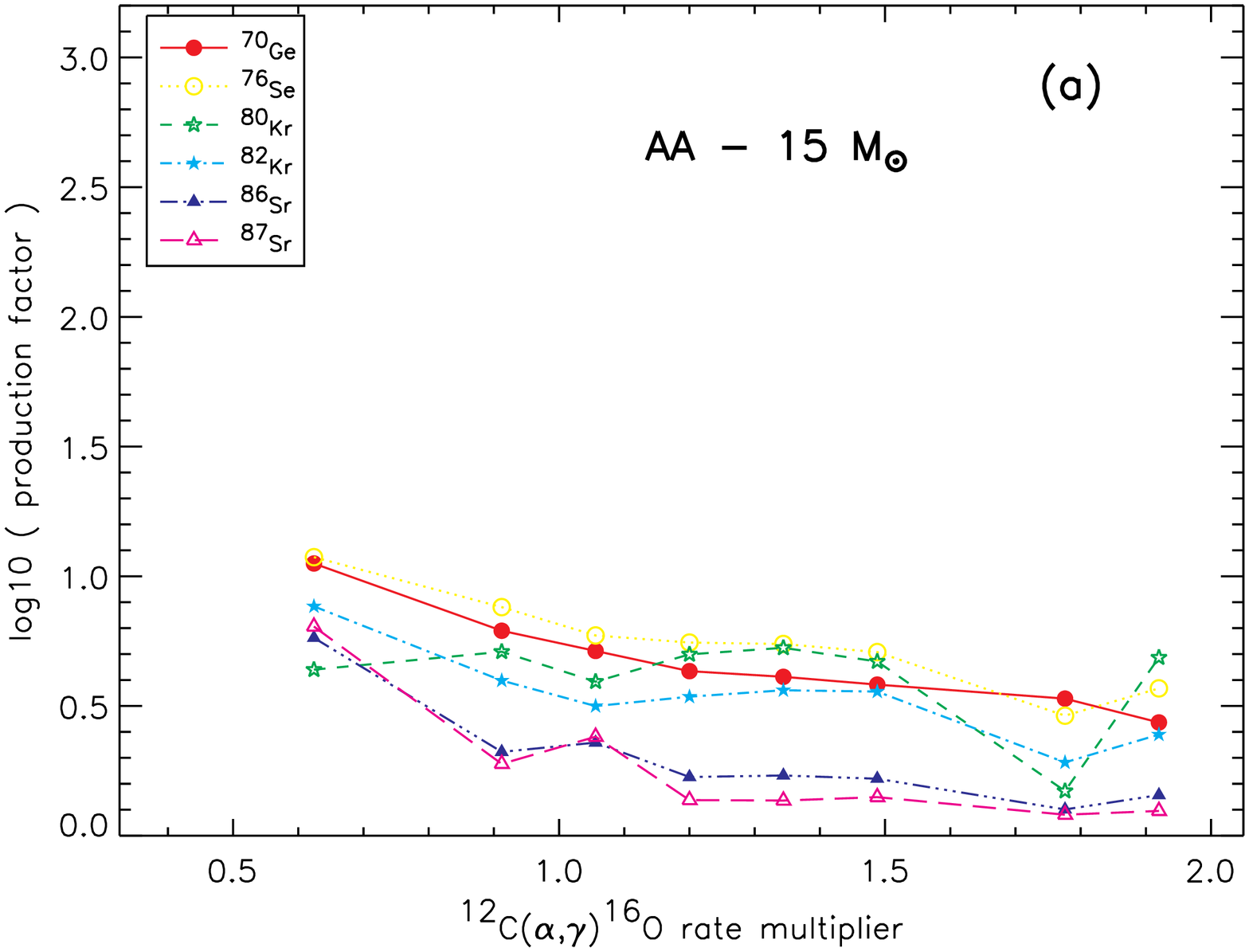}
\hfill
\includegraphics[angle=90,width=0.475\textwidth]{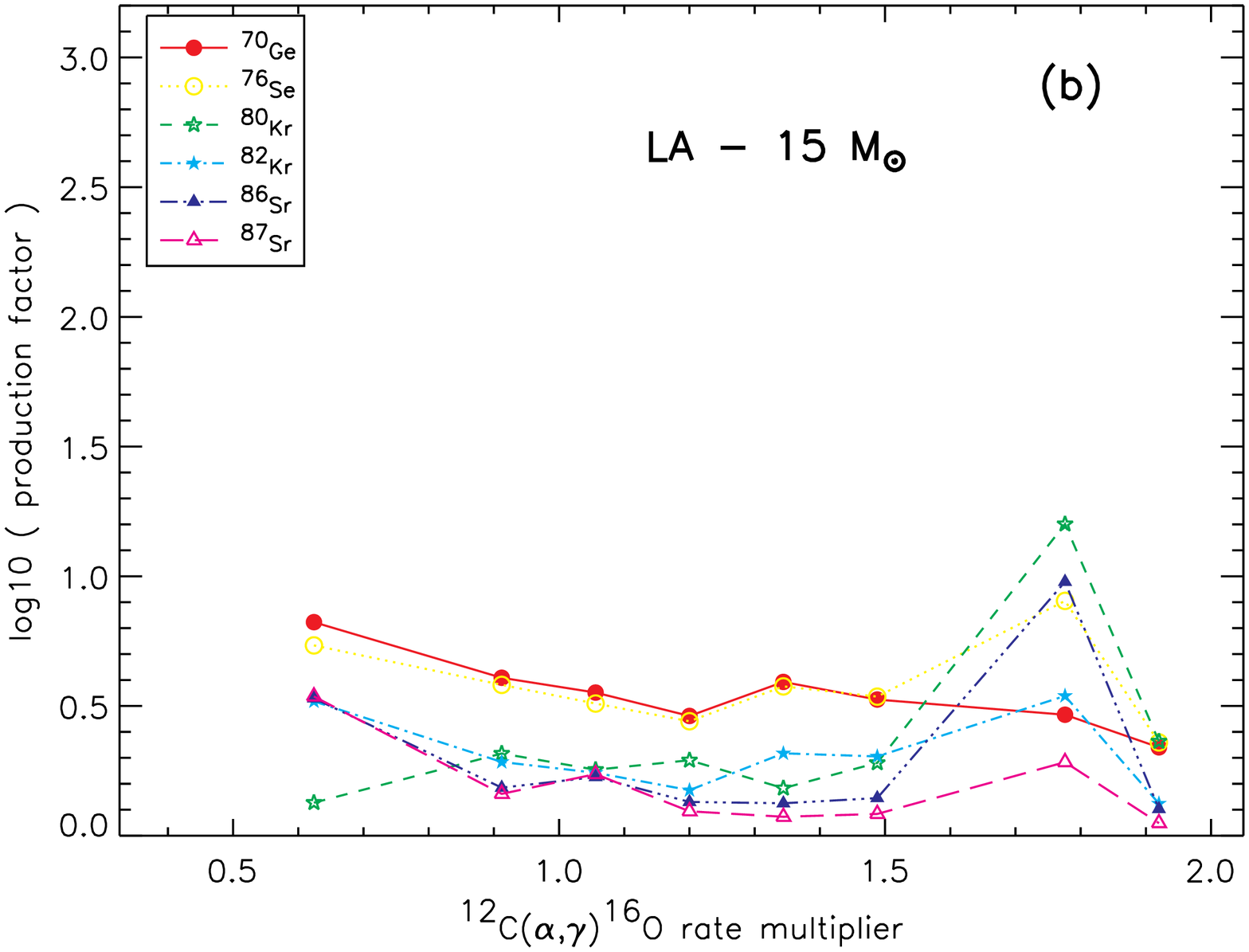}
\hfill
\includegraphics[angle=90,width=0.475\textwidth]{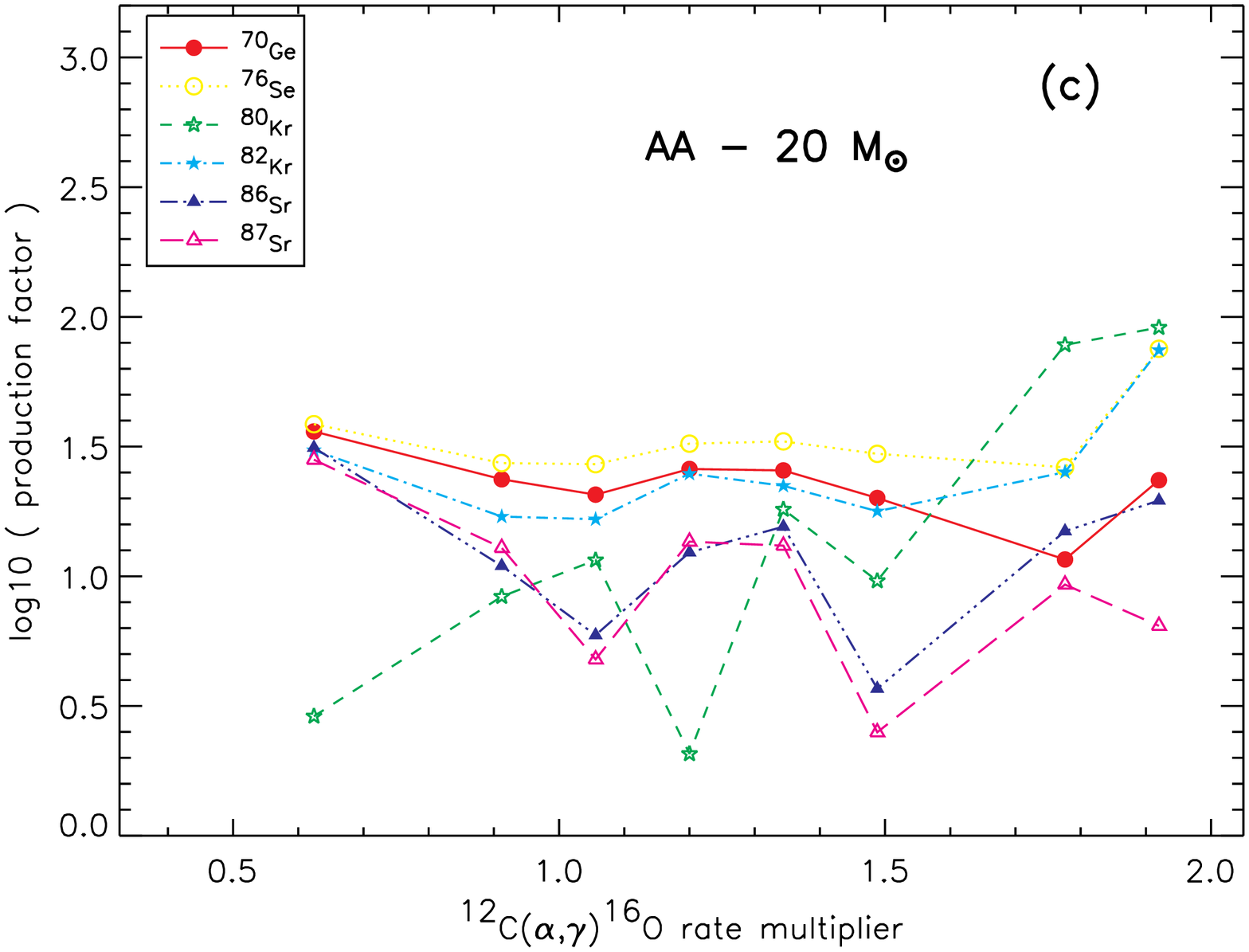}
\hfill
\includegraphics[angle=90,width=0.475\textwidth]{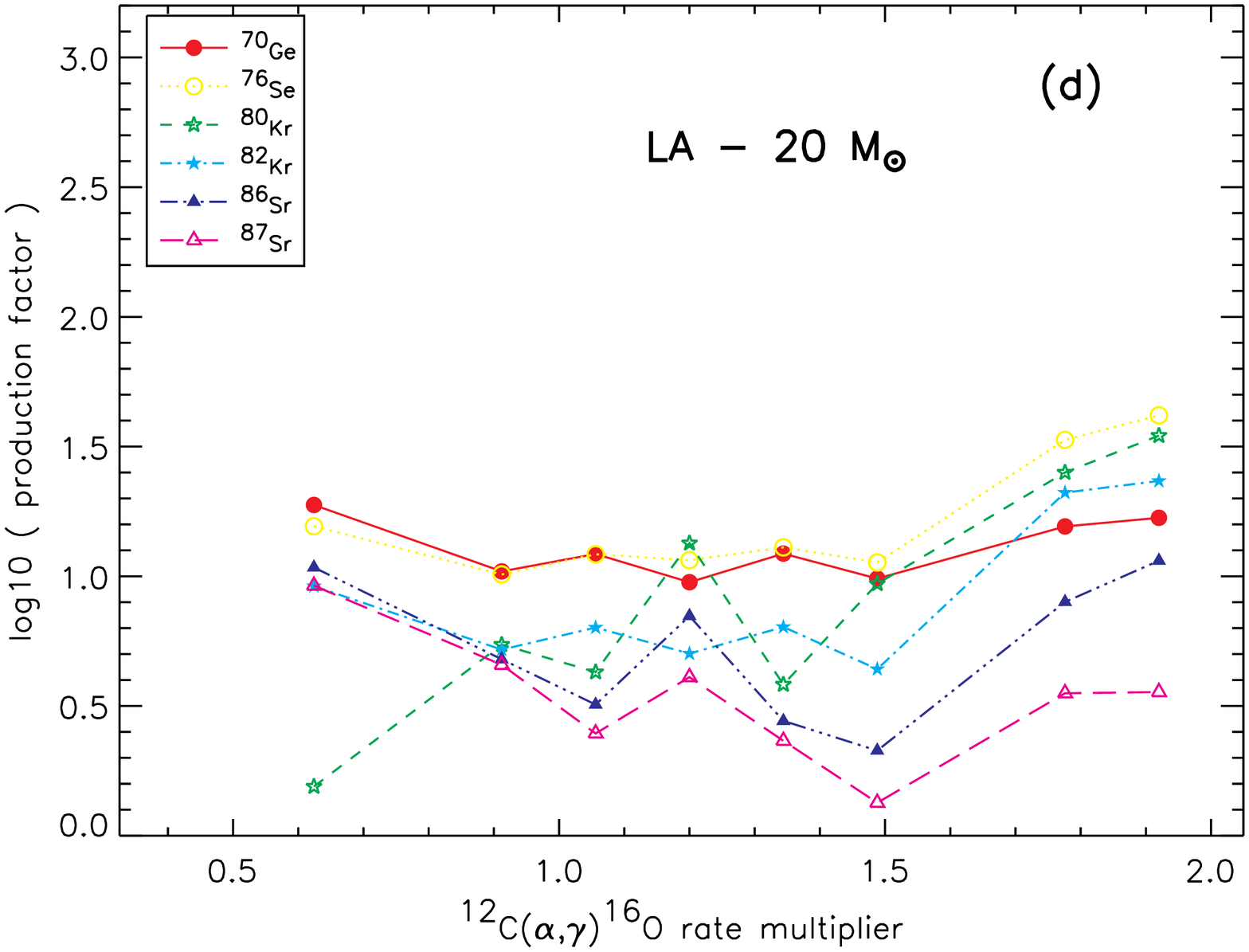}
\hfill
\includegraphics[angle=90,width=0.475\textwidth]{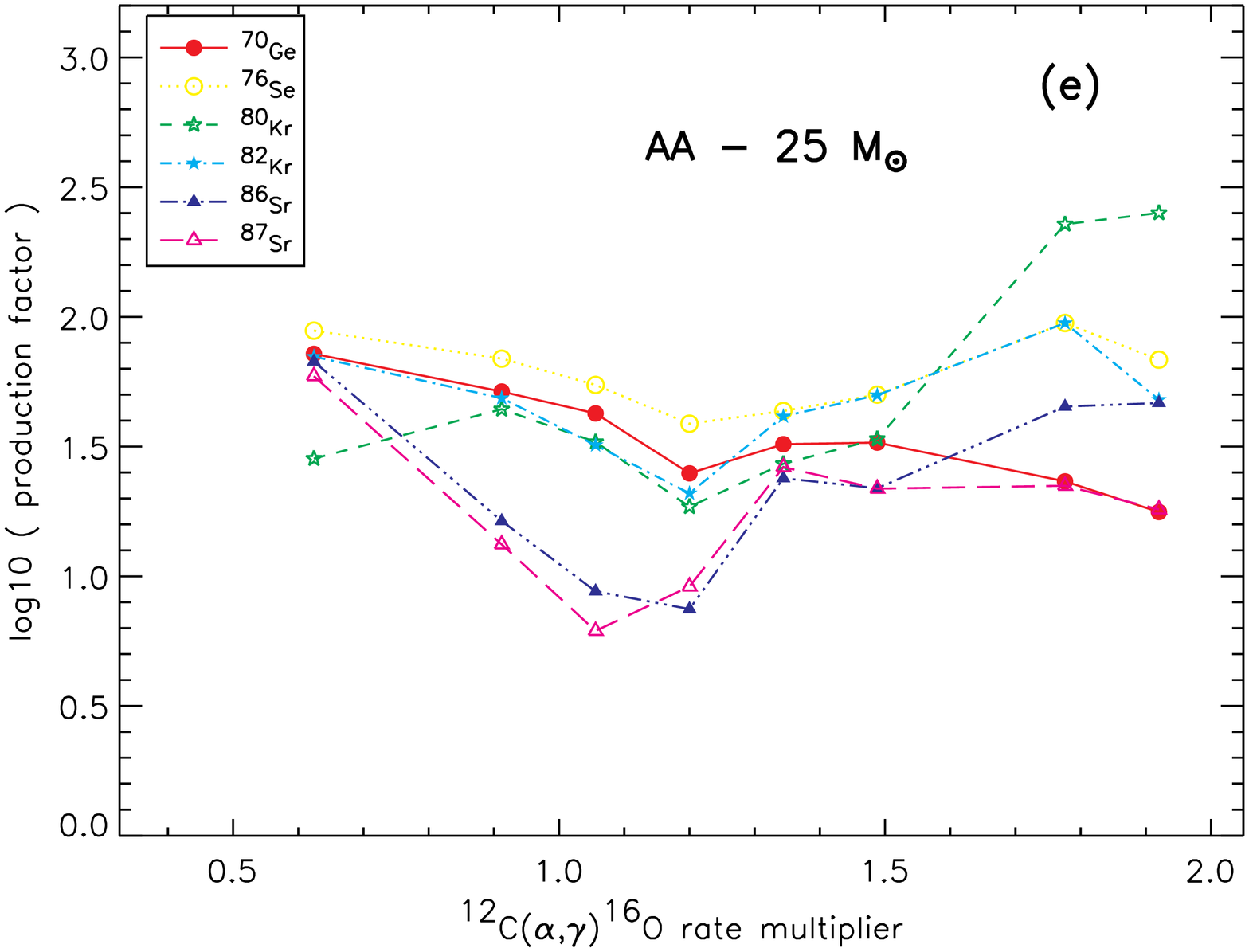}
\hfill
\includegraphics[angle=90,width=0.475\textwidth]{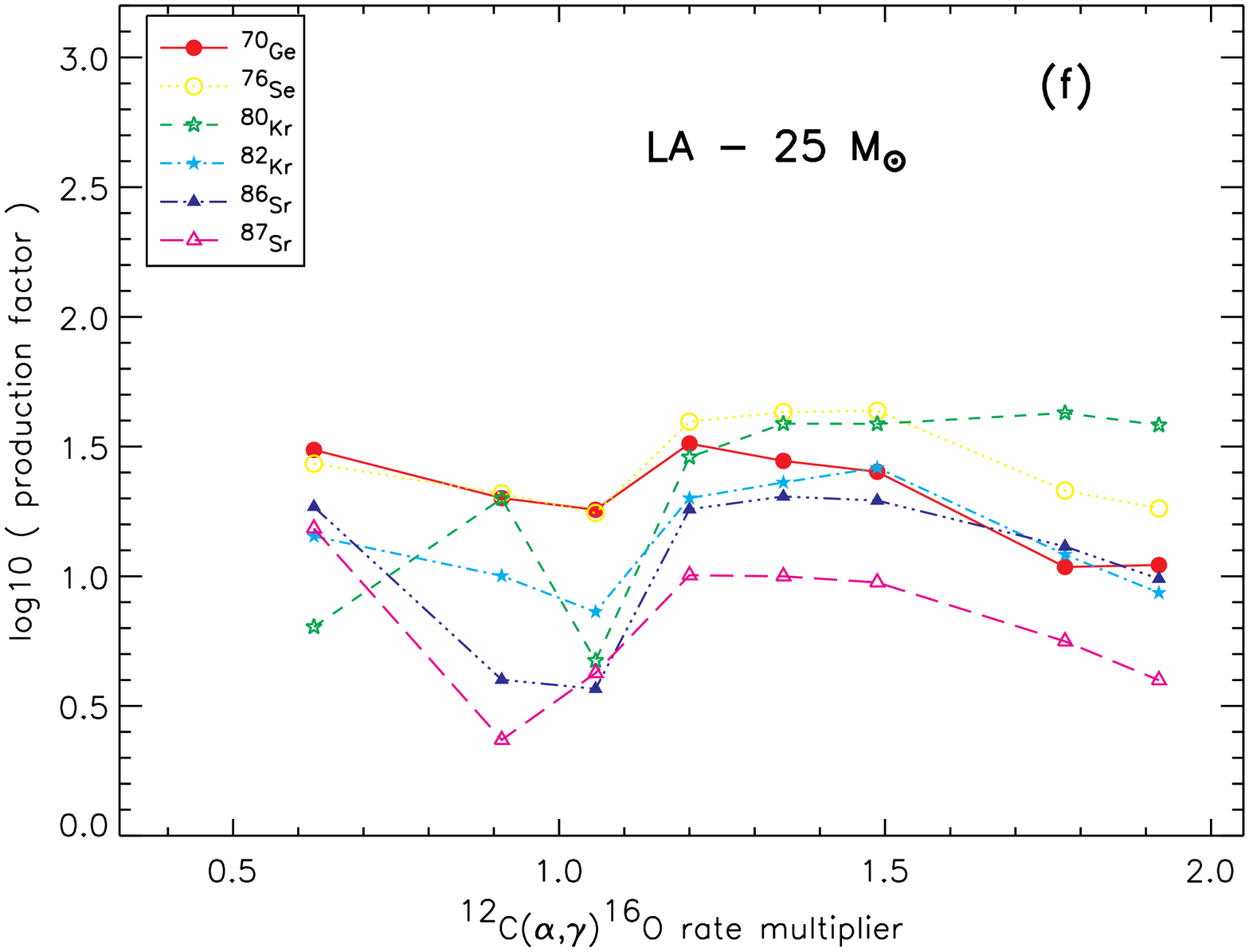}
\caption{Post-explosive production factors for s-only isotopes as a function of
  $R_{\alpha,12}$. \textbf{(a)} The AA series,
  $15\,\Msun$. \textbf{(b)} The LA series, $15\,\Msun$. \textbf{(c)}
  The AA series, $20\,\Msun$.  \textbf{(d)} The LA series,
  $20\,\Msun$. \textbf{(e)} The AA series, $25\,\Msun$. \textbf{(f)}
  The LA series, $25\,\Msun$.} \label{sOnlyPfVs12cag}
\end{figure*}

\begin{figure*}
\centering
\includegraphics[angle=0,width=\columnwidth]{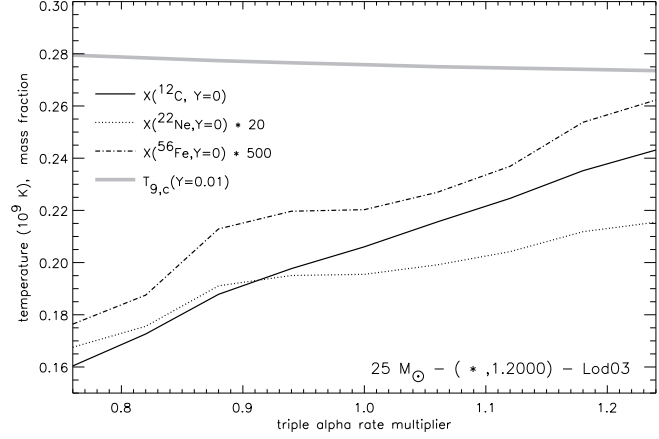}

\caption{Central temperature at the end of central helium burning
  ($1\,\%$ helium mass fraction - ``$Y=0.01$'') in units of $10^9\,$K
  ($T_{9,\mathrm{c}}$), central abundances of $^{12}$C, $^{22}$Ne, and
  $^{56}$Fe after core helium depletion (``$Y=0.00$'') for $25\,\Msun$
  stars with L03 initial abundance as a function of $R_{3\alpha}$.
}\label{hedep}
\end{figure*}

\begin{figure*}
\centering
\includegraphics[angle=90,width=0.475\textwidth]{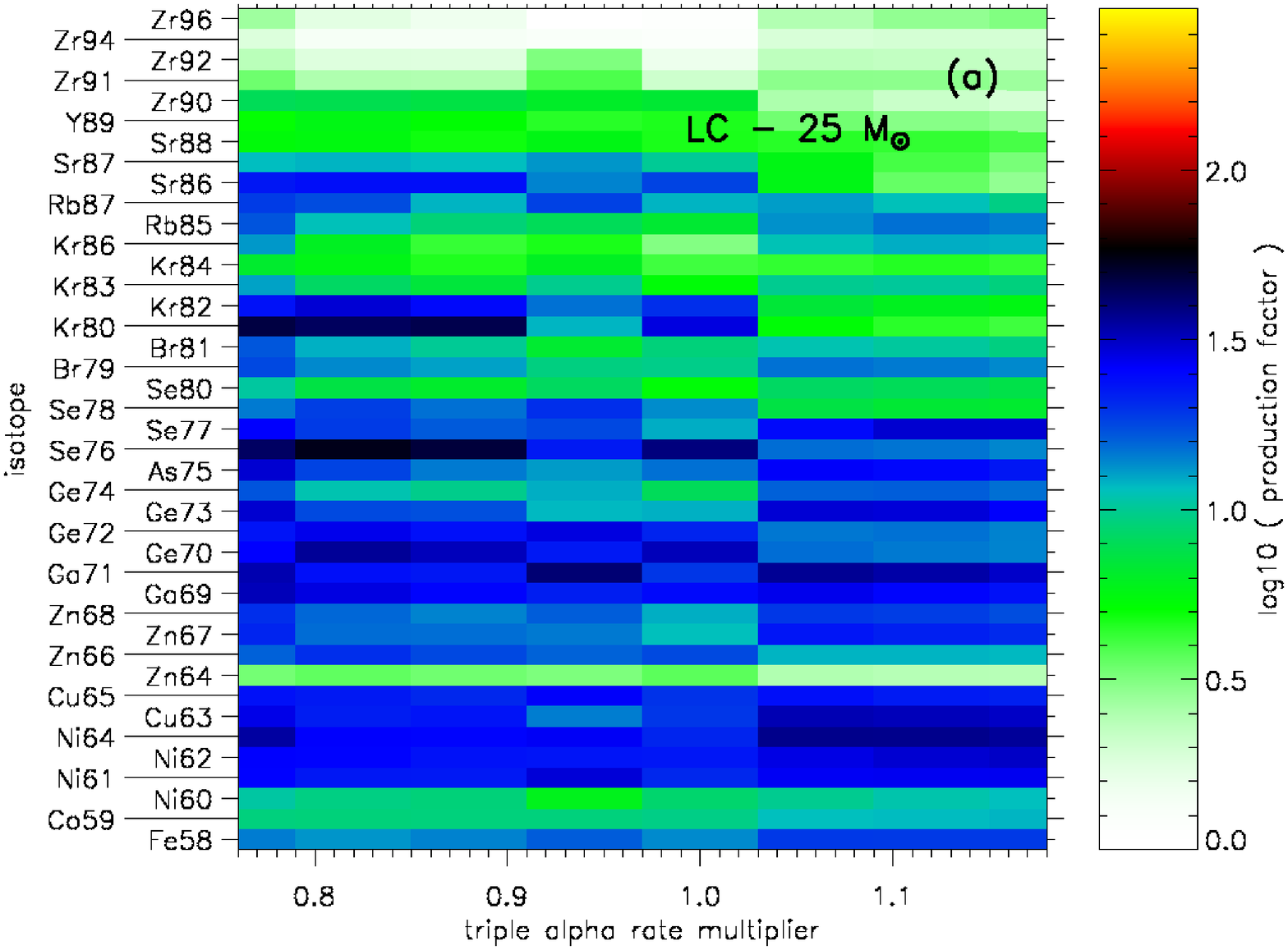}
\hfill
\includegraphics[angle=90,width=0.475\textwidth]{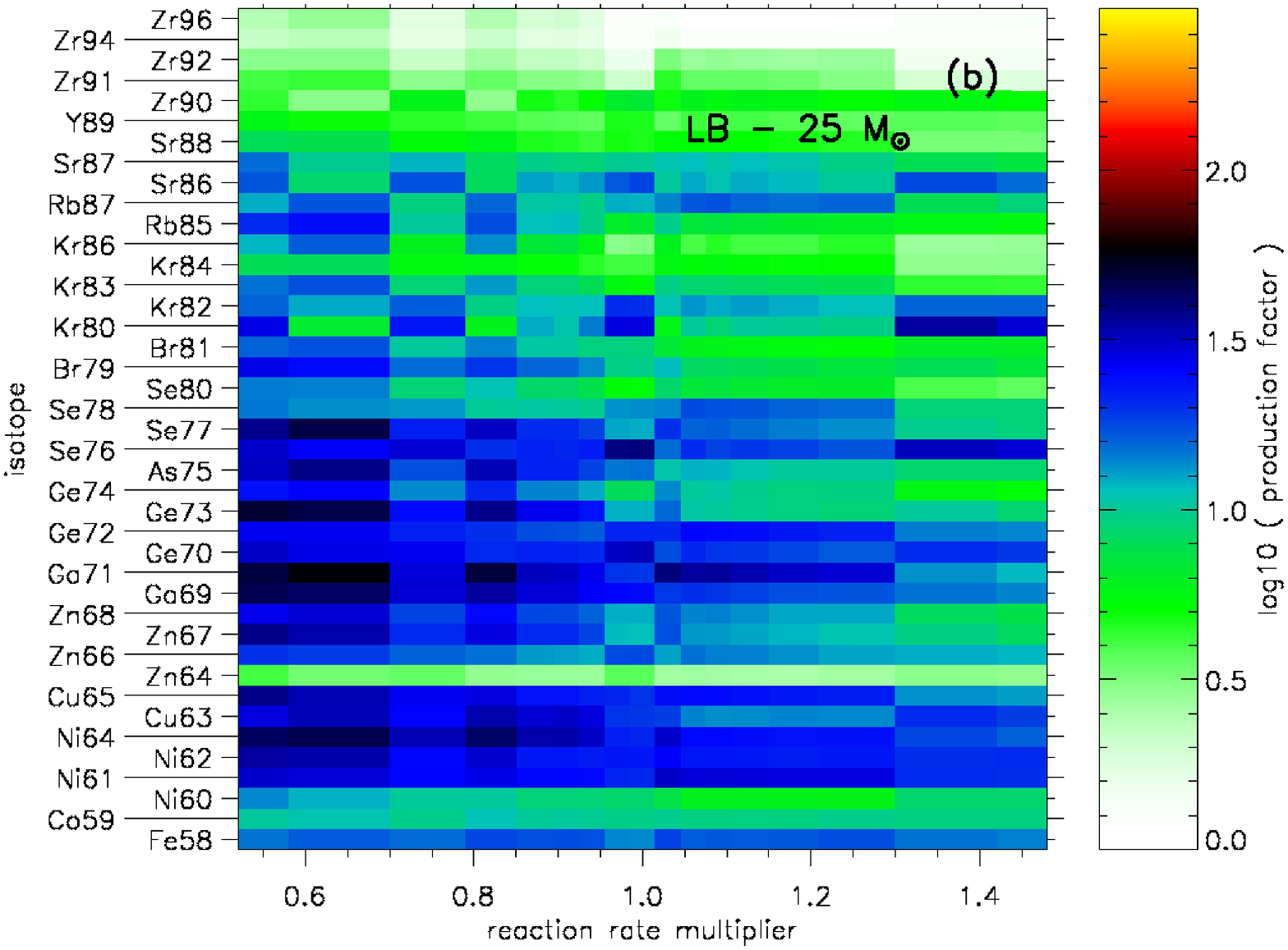}
\caption{Post-explosive production factors for a 25 \Msun star as a
  function of $R_{3\alpha}$. \textbf{(a)} The LC series. \textbf{(b)}
  The LB series.  The ``reaction rate multiplier'' is the factor that
  multiplies both the standard rates.}
 \label{pfVs3alpha}
\end{figure*}

\begin{figure*}
\centering
\includegraphics[angle=90,width=0.475\textwidth]{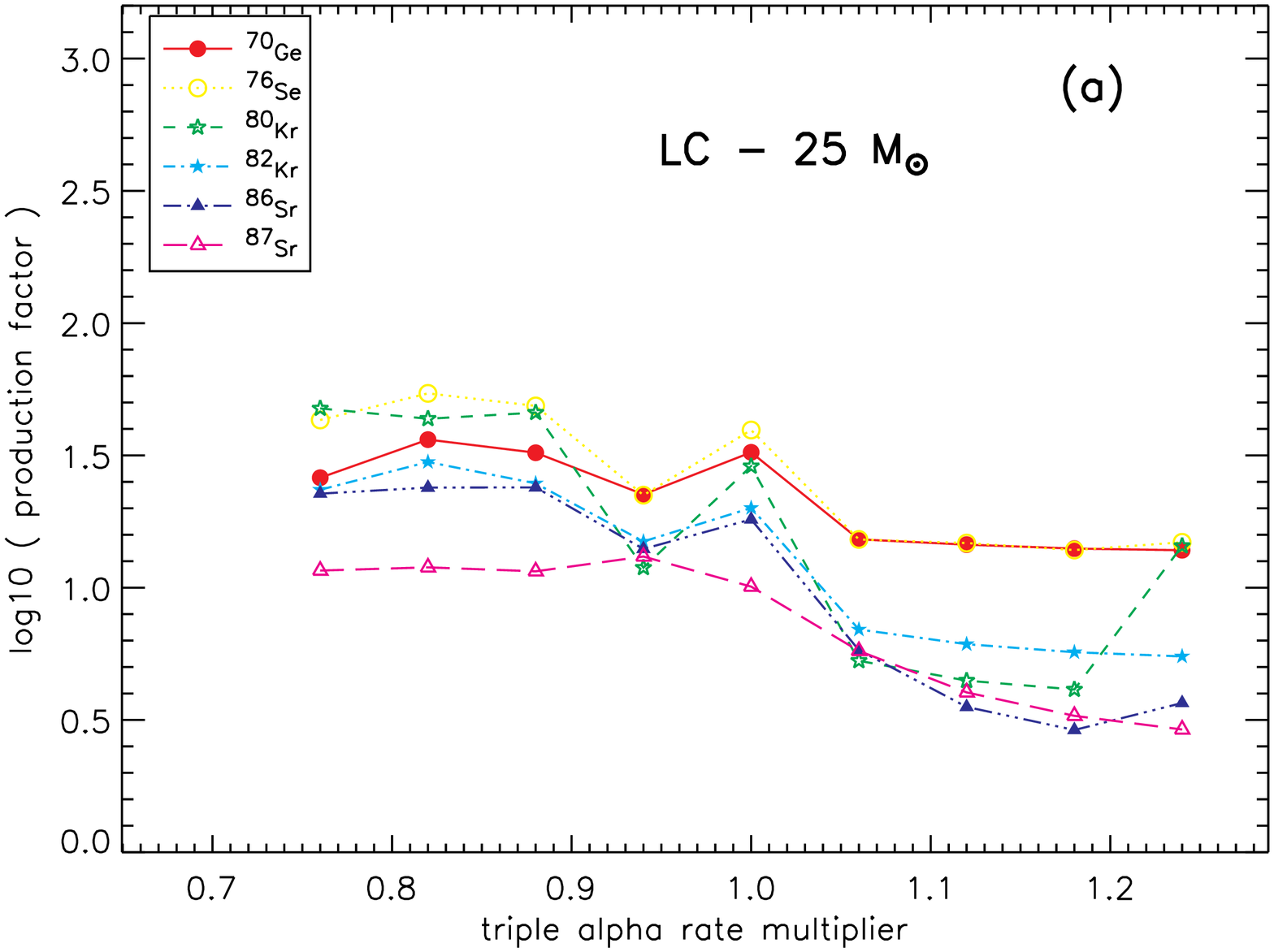}
\hfill
\includegraphics[angle=90,width=0.475\textwidth]{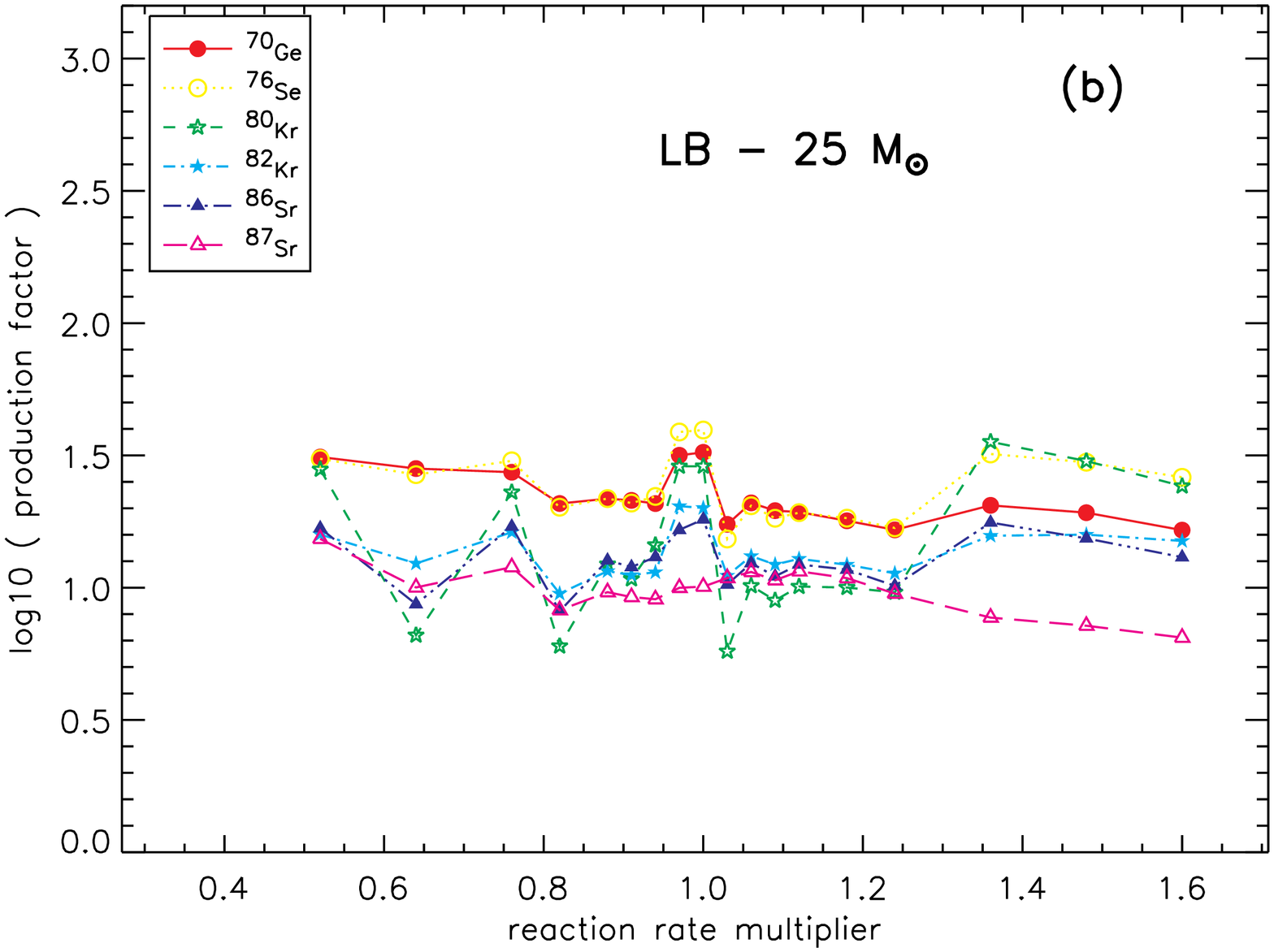}
\caption{Post-explosive production factors for s-only isotopes for a 25 \Msun 
star as a function of $R_{3\alpha}$. \textbf{(a)} The LC series. \textbf{(b)}
  The LB series.  The ``reaction rate multiplier'' is the factor that
  multiplies both the standard rates.}
 \label{sOnlyPfVs3alpha}
\end{figure*}

\section{Stellar Models and Input Physics}

The one-dimensional hydrodynamics code KEPLER
(\citealt{wea78}; \citealt{woo95}; \citealt{rau02}; \citealt{woo02}) 
is used to evolve stars from central H
burning up to Fe core-collapse.  The supernova explosion that follows
is simulated using a spherical piston placed at the base of the O
burning shell, which first moves inward and then outward at a constant
acceleration which has been adjusted to result in a total kinetic
energy of the ejecta of $1.2 \times 10^{51}$ ergs one year after the
explosion, and coming to rest at a radius of $1,000\,$km
\citep{rau02,woo07}.  After estimating the fallback from our
hydrodynamic supernova simulations, the final nucleosynthesis yields
are determined by employing the same parameterization of mixing as was
used by \citet{woo07}.

We performed calculations for stars of 15, 20, and $25\,\Msun$ and for
two different initial abundances, L03 and AG89.  Rate
sets were: (1) $R_{3\alpha}$ was kept constant (at its value from
\citealt{cau88}), and $R_{\alpha,12}$ was varied; (2) both rates were
varied by the same factor, so their ratio remained constant; and (3)
$R_{\alpha,12}$ was held constant at $1.2$ times the rate recommended
by \cite{buc96} and $R_{3\alpha}$ was varied.  Both reaction rates
were varied within a range of $\pm 2\sigma$ of their experimental
uncertainties.

For more details see \citet{tur07,woo07,rau02} who give a complete
description of the improvements to the stellar physics and reaction
rates since \cite{woo95}.  In particular, the ``rath'' rates, \cite{rau00},
have been adopted for the competing reactions
$^{22}$Ne($\alpha,n$)$^{25}$Mg and
$^{22}$Ne($\alpha,\gamma$)$^{26}$Mg.

We adopt a two-character notation to label our plots, e.g., LA, similar
to the notation adopted by \cite{tur07}. The first character can be an
L (to denote the L03 initial abundances) or an A (for the
AG89 initial abundances).  The second character denotes the
study: A, when $R_{3\alpha}$ was kept constant, and $R_{\alpha,12}$
was varied, B, when both rates were varied by the same factor, so
their ratio remained constant, and C, when $R_{\alpha,12}$ was held
constant, and $R_{3\alpha}$ was varied.

\section{Post-Explosive Production Factors}

In Figure~\ref{pfVs12cag}, we show the evolution of the production
factors as a function of $R_{\alpha,12}$ for isotopes between
$^{58}$Fe and $^{96}$Zr.  To facilitate reading, Figure \ref{sOnlyPfVs12cag}
shows the same for the 6 s-only isotopes. We define the production factor 
for each stable isotope as the ratio of the mass fraction of the isotope in the
supernova ejecta to its initial solar mass fraction.  $R_{3\alpha}$
was kept constant at the value from \cite{cau88}.  The color bar range
is from $1$ to $3 \times 10^{2}$.  The range of production factors for
a $25\,\Msun$ star are shown in Table \ref{minmax}. For most assumed reaction
rates, production factors (PF) obtained using the L03
abundances (Figures \ref{pfVs12cag}\emph{b}, \ref{pfVs12cag}\emph{d},
and \ref{pfVs12cag}\emph{f}), are significantly smaller than those for
the AG89 abundances (Figures \ref{pfVs12cag}\emph{a},
\ref{pfVs12cag}\emph{c}, and \ref{pfVs12cag}\emph{e}). Presumably the
lower CNO content of the L03 abundance set leads to a smaller
amount of $^{22}$Ne, lower neutron abundance, and a less efficient
\textsl{s}-process \citep{woo07}.

We have shown in \cite{tur07} that for the range of reaction rates we
calculate, the PF for $^{16}$O is $10.0 \pm 0.5$ for the
AG89 abundances, and $15.3 \pm 0.5 $ for the L03
abundances; the uncertainty is the standard deviation of the results
for the different rates. We find that the PFs we obtain for the
\textsl{s}-process elements are often larger than this: see Figure 1
and Table \ref{minmax}. This is not surprising: very low metallicity stars
produce oxygen, but very small abundances of the secondary weak
\textsl{s}-process nuclei.  We next use a simple model to estimate
what PF in a solar-abundance star would be required to reproduce the
observed abundances.

Assume that a nuclide $X$ is to be produced entirely by secondary
processes in massive stars.  It follows that when integrated over
metallicities from zero to the present, its PF must be the same as
that for $^{16}$O. Solar metallicity stars must, therefore, have a
larger PF for $X$ than for $^{16}$O. In a simple closed-box model,
assuming the production of $X$ is proportional to metallicity, the
required ratio of secondary to primary production in a solar abundance
star is two.  We take this as a working hypothesis, bearing in mind
that it is only a rough approximation.  Then, if the PF for a
weak-\textsl{s} nuclide is $2\times$PF($^{16}$O), massive stars
produce the observed solar abundance, or if it is equal to
PF($^{16}$O), half the observed abundance.  The fraction $f$ of the
observed solar abundance for nuclide $X$ produced in our model is
given by
\begin{equation}
f(X) =  0.5\times\textrm{PF}(X)/\textrm{PF}(^{16}\textrm{O})
\label{PFeq}
\end{equation}
where the $0.5$ arises from the model described above; as noted above
PF($^{16}$O) is approximately 10 (15.3) for the AG89 (L03) abundances.
Equation ~(\ref{PFeq}) and Figure ~\ref{pfVs12cag} can be used to estimate
whether a particular isotope is overproduced for a given star or rate
choice.

The production factors increase strongly as the stellar mass
increases, reflecting a more efficient \textsl{s}-process as the
temperature increases. In particular, most isotopes are weakly
produced, PF$<$20 (30) for AG89 (L03) in the $15\,\Msun$ star, with
the exception of those between $^{58}$Fe and $^{65}$Cu, and of some
isotopes when $R_{\alpha,12}$ is given its lowest or highest value
(Figures \ref{pfVs12cag}\emph{a} and \ref{pfVs12cag}\emph{b}). The
highest PF are observed for the $25\,\Msun$ star (Figures
\ref{pfVs12cag}\emph{e} and \ref{pfVs12cag}\emph{f}), whereas the
$20\,\Msun$ star shows a contribution between those two extremes
(Figures~\ref{pfVs12cag}\emph{c} and \ref{pfVs12cag}\emph{d}).
Ignoring the contribution of the $15\,\Msun$ and $20\,\Msun$ stars as
was done in \cite{rai91a} and \cite{rai91b} does not seem justified;
their contributions are not always negligible.

We find that $^{62}$Ni is overproduced for many rate choices,
especially for the $25\,\Msun$ star and the AG89 abundances.
This problem was already documented by \cite{rau02} and seems to arise
from the decreased value of the $^{62}$Ni($n,\gamma)^{63}$Ni reaction
rate used for our models compared to its previous compilation.  The
factor of three change arises from different extrapolations of a cross
section measured only at thermal neutron energies (see also
\citealt{nas05}, however).

Table~\ref{minmax} shows, for a $25\,\Msun$ star of L03
initial abundance, the value of the maximum and minimum production
factors for nuclides between $^{58}$Fe and $^{96}$Zr when either
$R_{\alpha,12}$ or $R_{3\alpha}$ is varied within its $\pm 2\sigma$
uncertainties.  Large differences in PFs are found when either
$R_{3\alpha}$ or $R_{\alpha,12}$ are varied.  Most commonly, the largest
PFs are obtained for the lower values of $R_{\alpha,12}$ as one sees 
in Figure~\ref{pfVs12cag}.  Figure~\ref{hedep} shows some of the changes 
resulting from changes in $R_{3\alpha}$. Lower values of $R_{3\alpha}$ 
lead to lower $^{12}$C abundance at the end of central
helium burning.  This, in turn, is compensated by a slightly higher
central temperature to produce the same energy release rate to
maintain the star's luminosity.  Due to the high
temperature sensitivity of the $^{22}$Ne$(\alpha$,n$)$ reaction, this increased
temperature means more
of the $^{22}$Ne has been burnt by central helium depletion, resulting in a
stronger \textsl{s}-process. Figure~\ref{hedep} also shows
the decrease of $^{56}$Fe for lower $R_{3\alpha}$ that correlates
very well with the smaller amount of $^{22}$Ne left, i.e., more of it
being burnt, providing a measure of neutron exposure.

For many isotopes, the maximum PF is greater than twice the
minimum. For example, for $\pm 2\sigma$ uncertainties in
$R_{\alpha,12}$ the PF for $^{73}$Ge ranges from 1.5 to 6.5; 4.2 to
20.9; 12.1 to 38.0 for $15\,\Msun$; $20\,\Msun$; $25\,\Msun$ stars,
resp. The \textsl{s}-only isotopes show a similarly strong
sensitivity.  For $\pm 2\sigma$ uncertainties in $R_{\alpha,12}$, the
PF for $^{70}$Ge, an \textsl{s}-only isotope, decreases from 6.6 to
2.2; 18.8 to 9.5; 32.5 to 10.9 for $15\,\Msun$; $20\,\Msun$;
$25\,\Msun$ stars.  For an average over the 3 stars, the
production factor for $^{80}$Kr, another \textsl{s}-only isotope, ranges
from 2.5 at the lowest value of $R_{\alpha,12}$ to 20.5 at its next
to highest value.

\begin{figure}
\centering
\includegraphics[angle=90,width=0.475\textwidth]{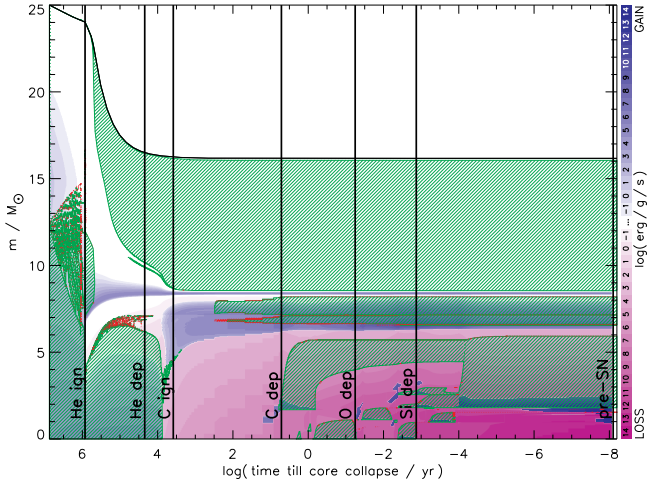}
\caption{Energy history of the $25\,\Msun$ star (central values of the
  reaction rates, L03 abundances) as a function of time until core
  collapse.  The ordinate is the included mass from the stellar
  center. The green cross hatched areas are fully convective, and the
  red cross hatched areas are semiconvective.  The blue and pink
  shading indicate net energy generation, with blue positive and pink
  negative.  For more details, see \cite{woo02}. The vertical lines
  show the times when dump files are generated by the KEPLER code at
  key stages of the evolutionary process.  }\label{conv}
\end{figure}

In Figure~\ref{pfVs3alpha}, we illustrate the variations in the
production factors for a $25\,\Msun$ star with the L03
initial abundances when $R_{3\alpha}$ is varied and $R_{\alpha,12}$ is
held constant (Figure \ref{pfVs3alpha}\emph{a}), and when both
reaction rates are varied by the same amount (Figure
\ref{pfVs3alpha}\emph{b}). To facilitate reading, we show the same for the 
6 s-only isotopes in Figure \ref{sOnlyPfVs3alpha}. Large variations in the 
production factors are observed in both cases, demonstrating the importance 
of both helium burning reactions.

The results shown in Figure~\ref{pfVs12cag} and
Figure~\ref{pfVs3alpha} are consistent with the findings of
\cite{tur07} that both reactions sensitively influence the production
factors of the medium-weight isotopes, among which are the neutron
poisons mentioned earlier.  When, however, we examine the production
of these poisons and of the weak-\textsl{s} isotopes, as a function of
the rates, we find only weak indications of any correlation.

\begin{figure*}
\centering
\includegraphics[angle=90,width=0.425\textwidth]{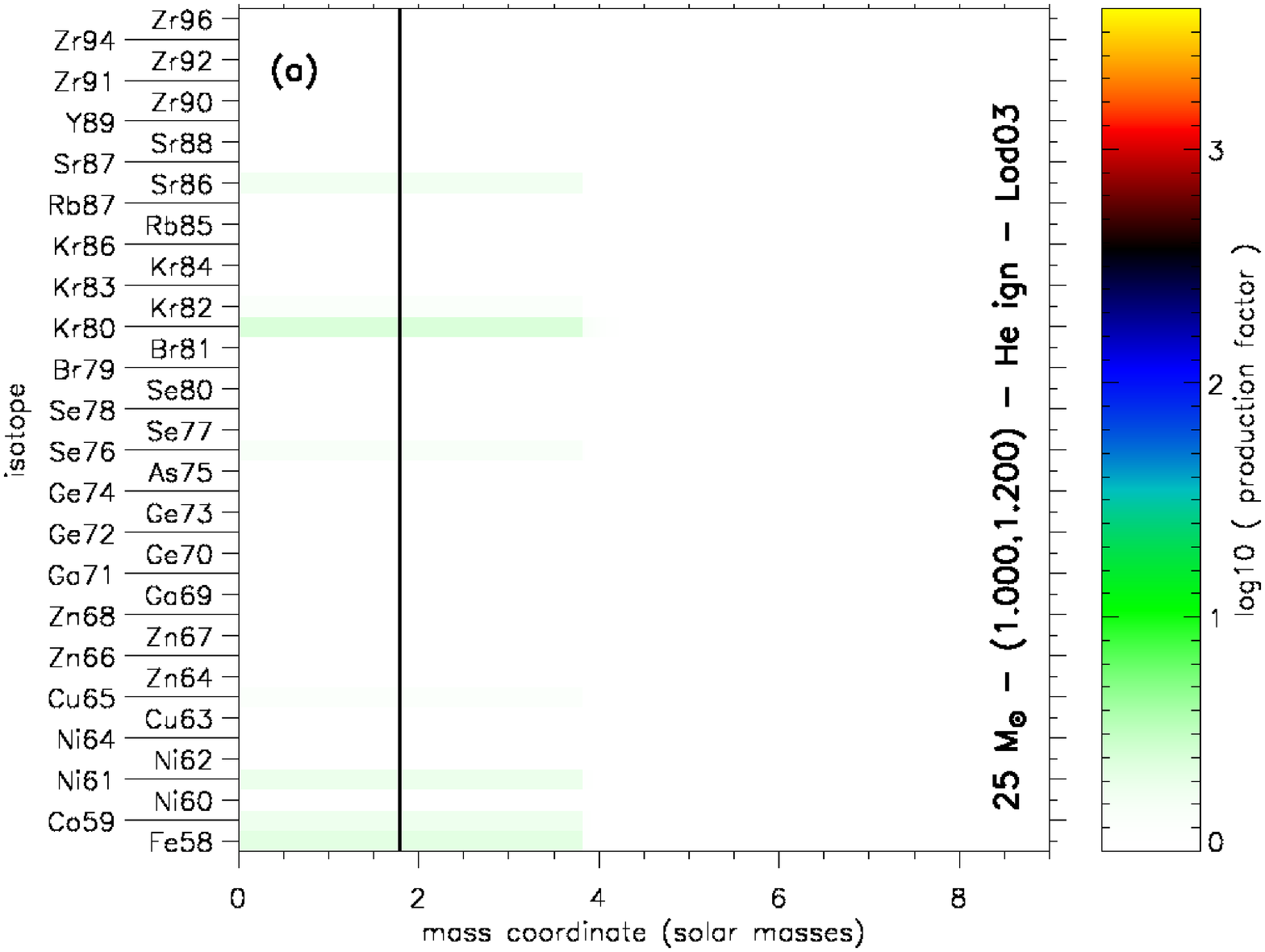}
\hfill
\includegraphics[angle=90,width=0.425\textwidth]{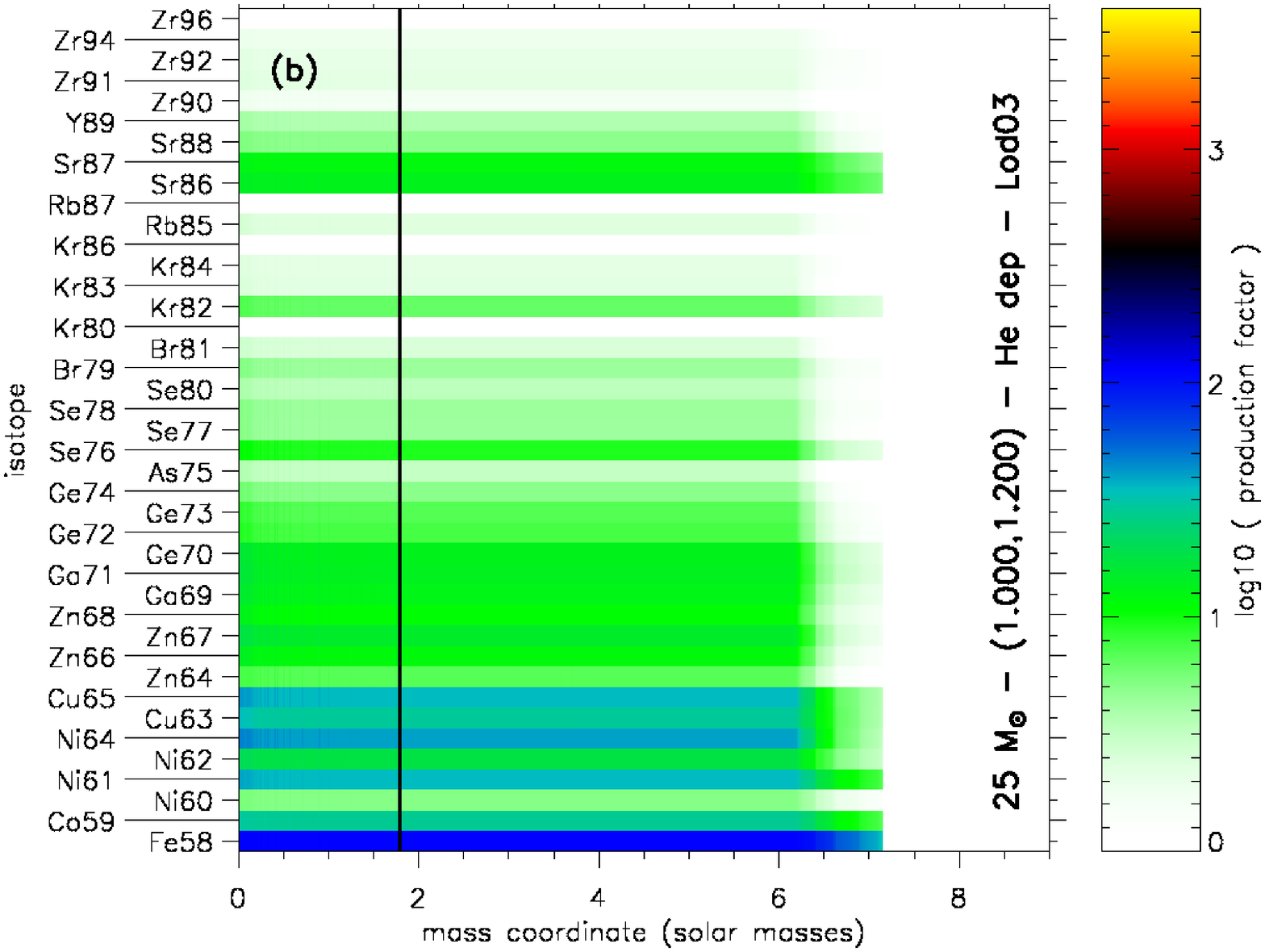}
\hfill
\includegraphics[angle=90,width=0.425\textwidth]{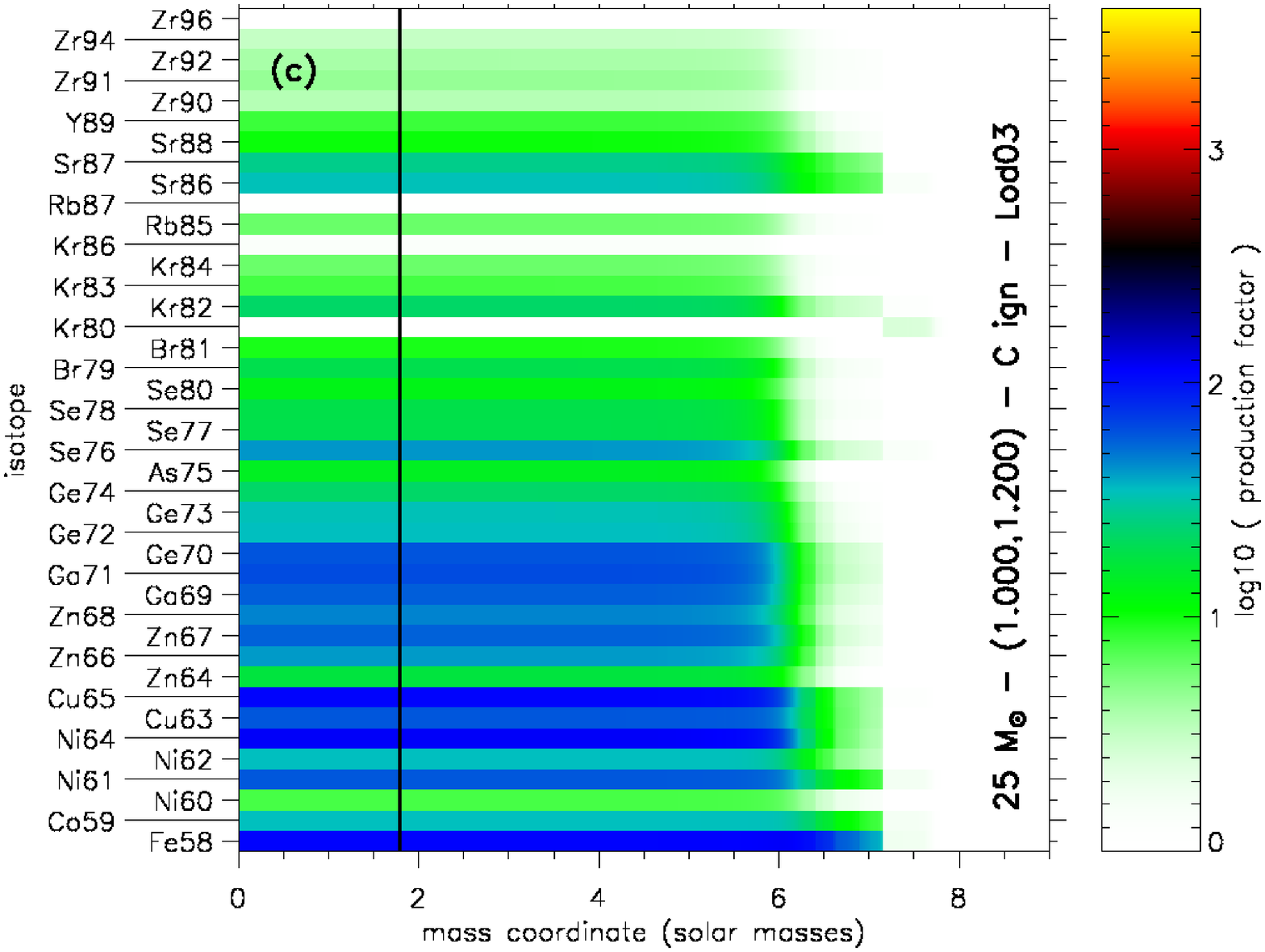}
\hfill
\includegraphics[angle=90,width=0.425\textwidth]{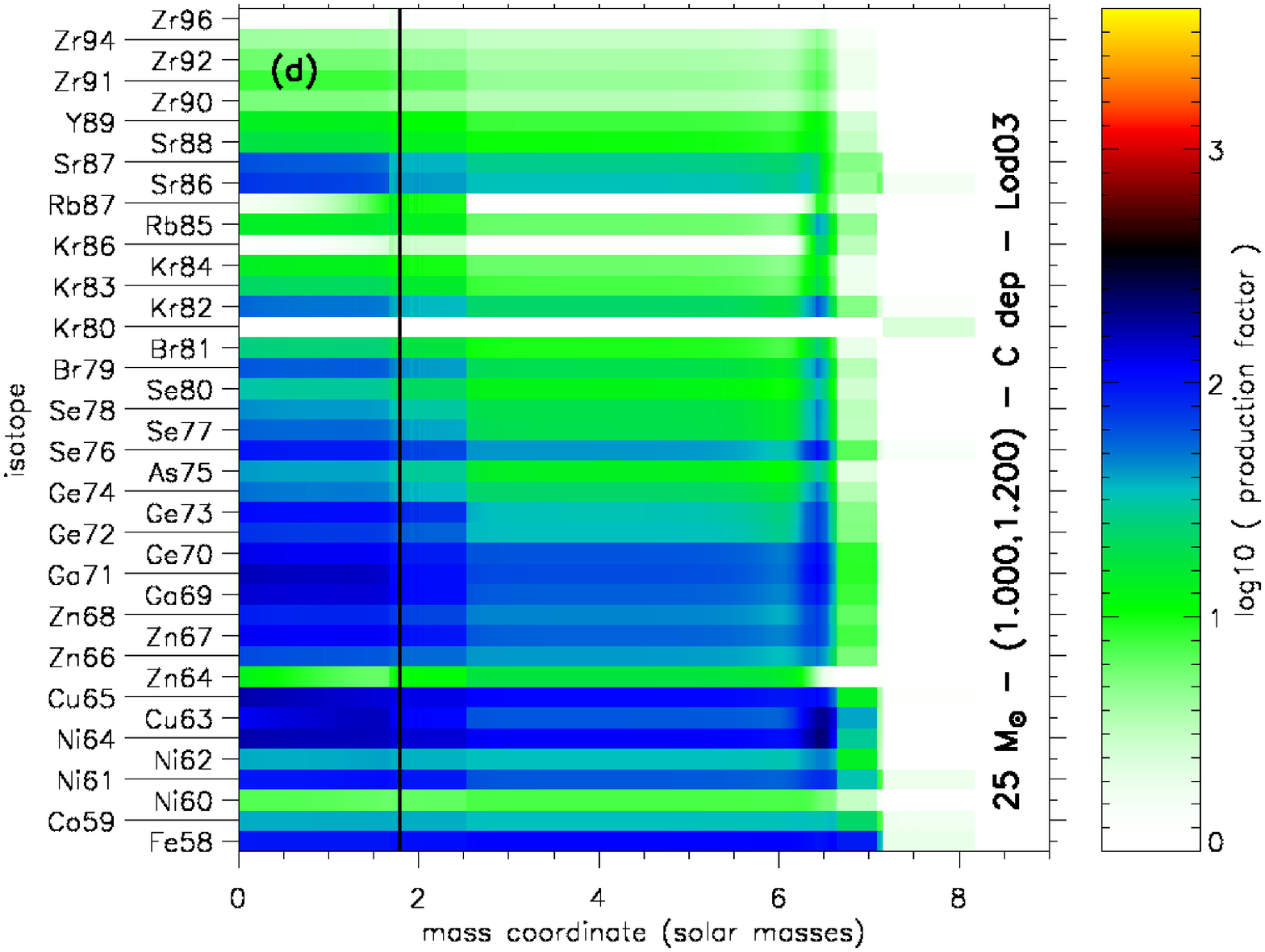}
\hfill
\includegraphics[angle=90,width=0.425\textwidth]{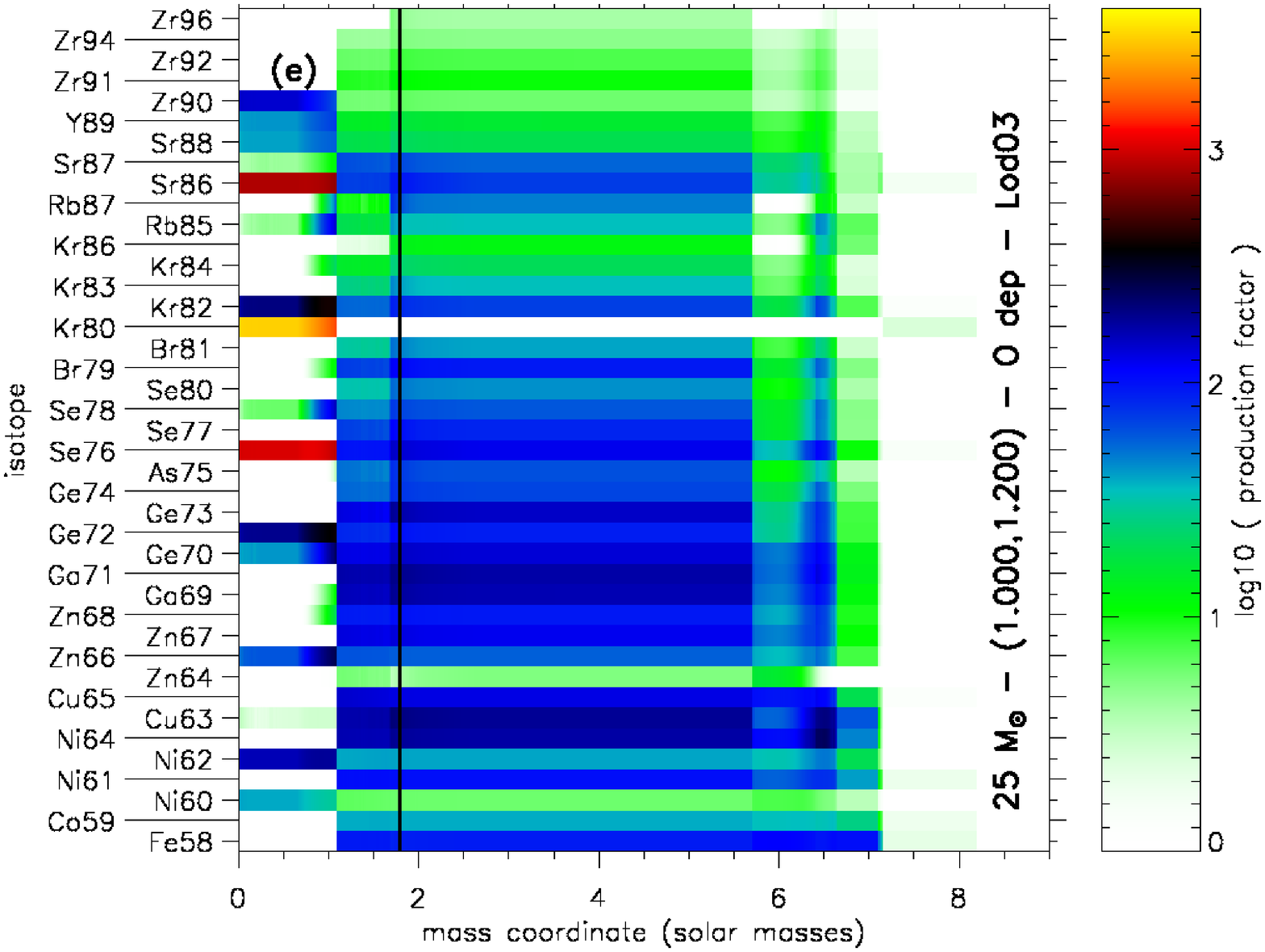}
\hfill
\includegraphics[angle=90,width=0.425\textwidth]{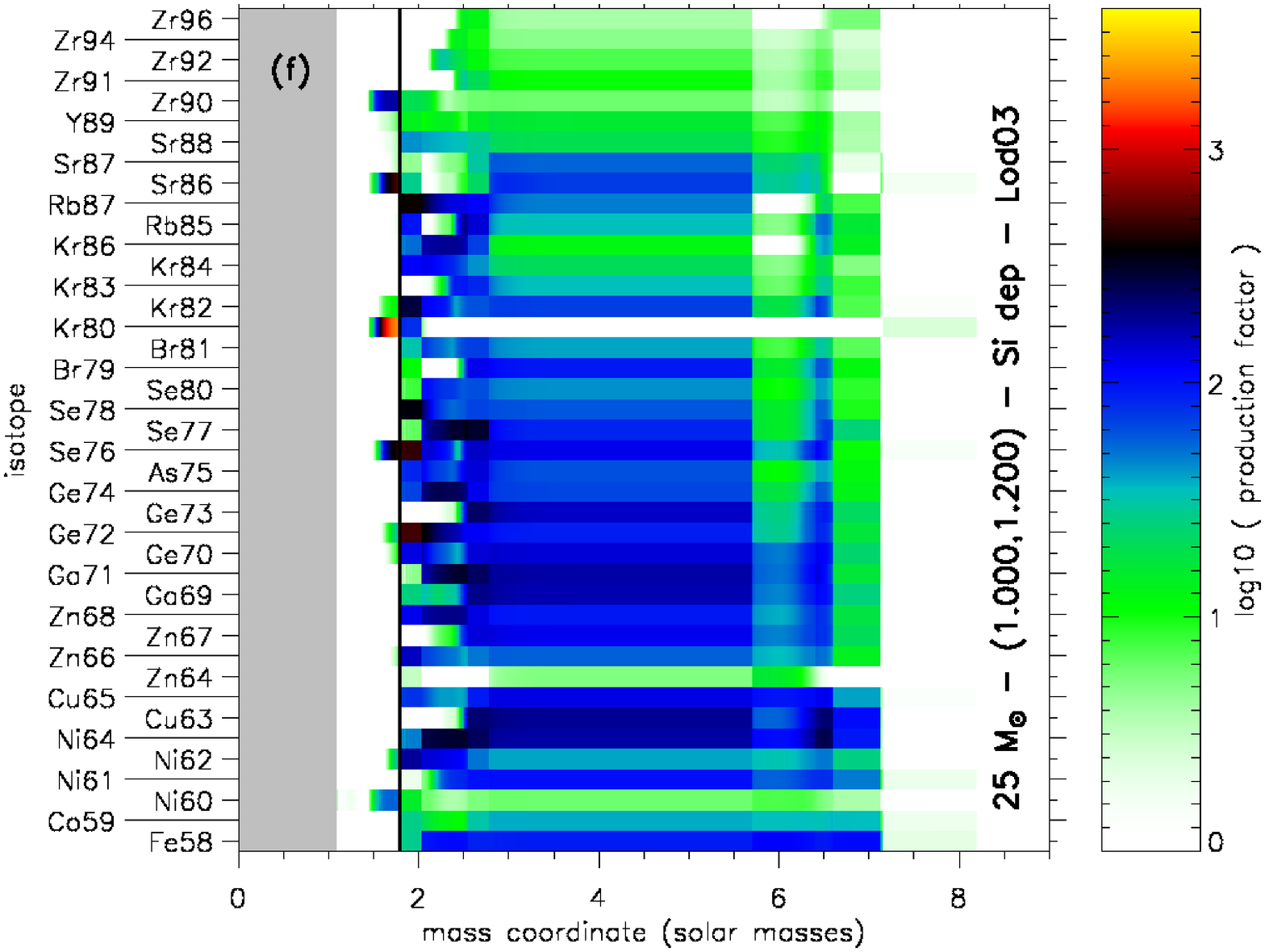}
\hfill
\includegraphics[angle=90,width=0.425\textwidth]{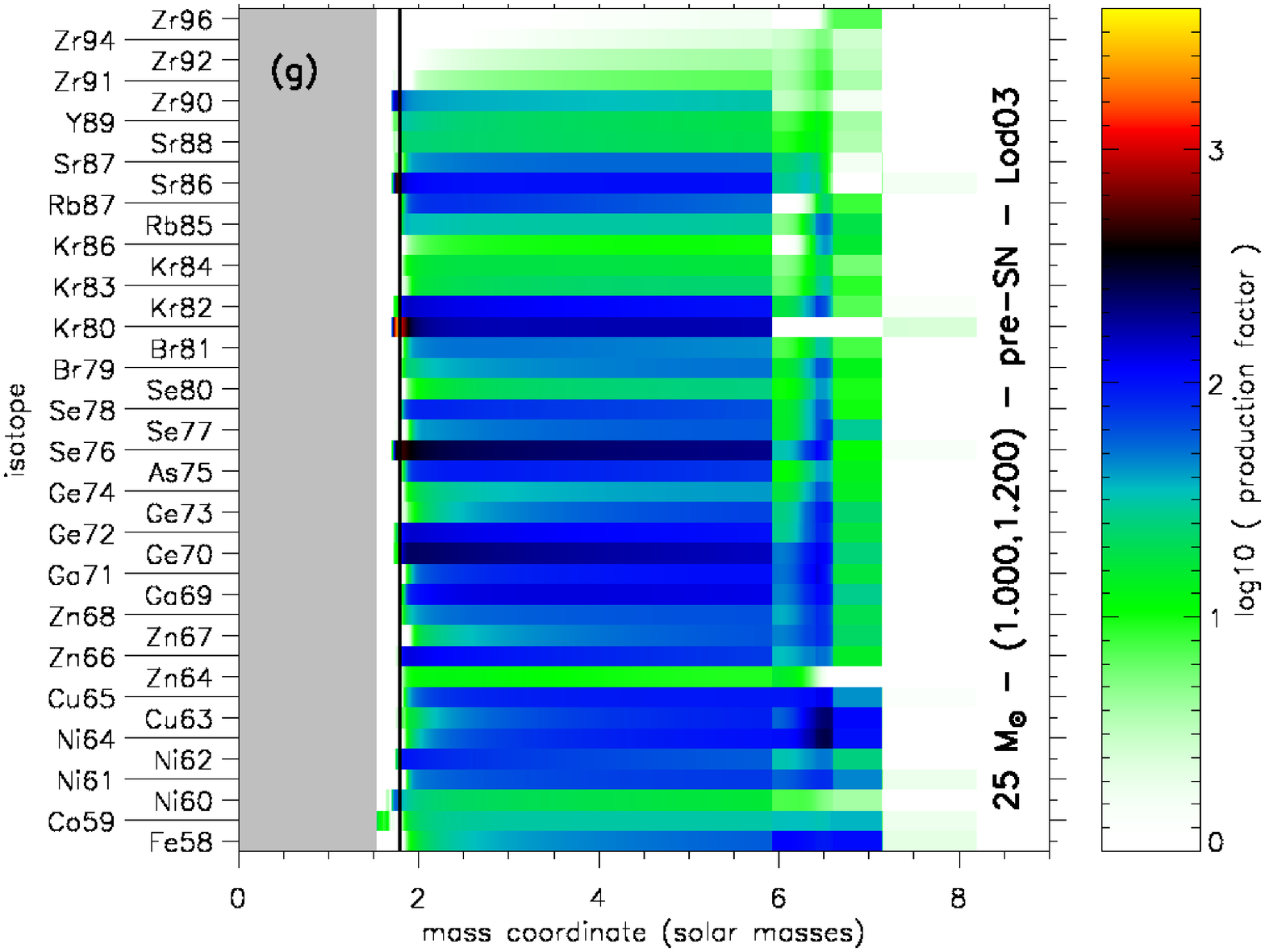}
\hfill
\includegraphics[angle=90,width=0.425\textwidth]{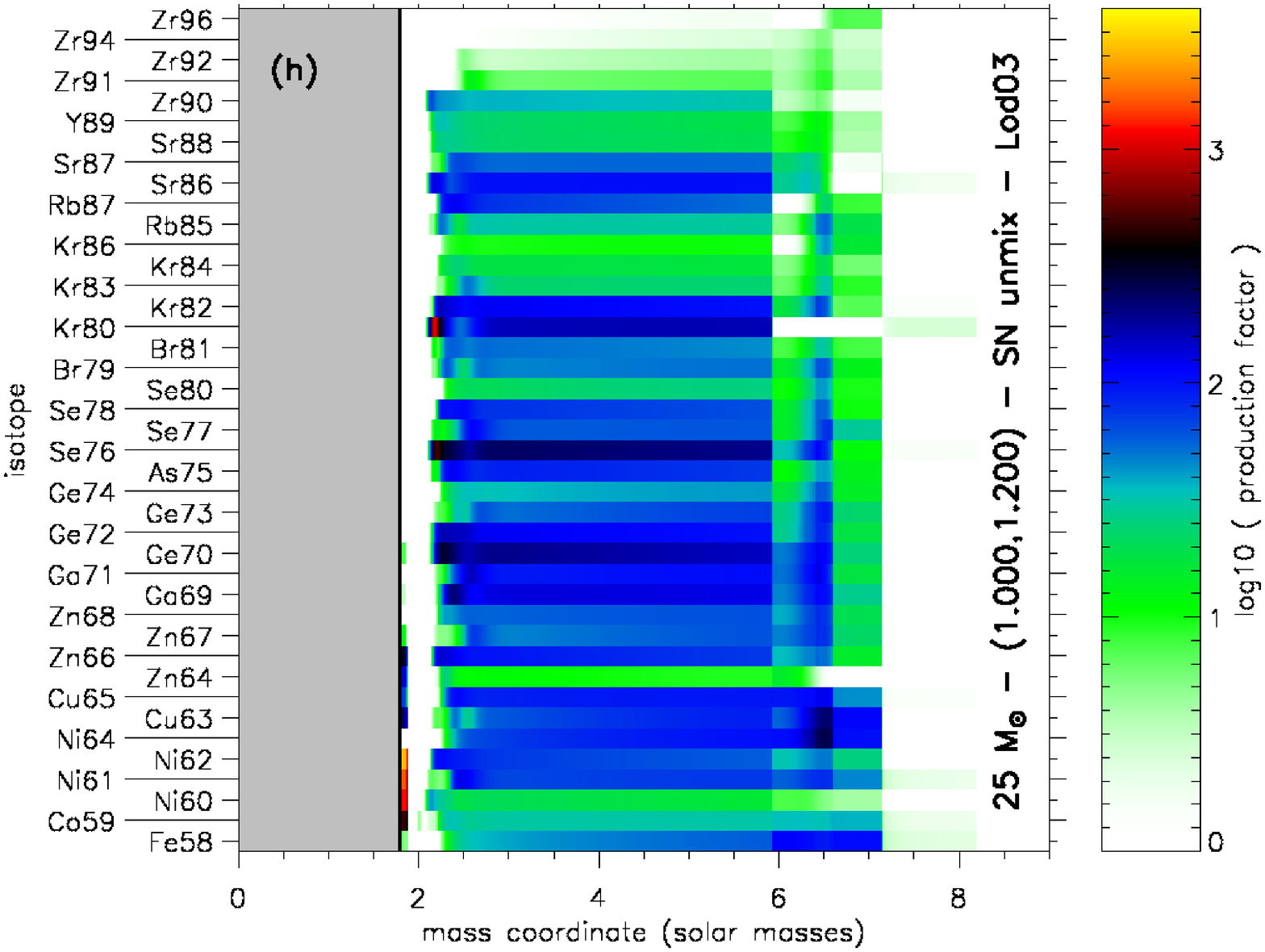}
\caption{Production factors for \textsl{s}-process isotopes for a $25\,\Msun$
star (L03 initial abundances) as a function of their
mass location within the star. \textbf{(a)} At He ignition.
\textbf{(b)} At He depletion (1\% He left). \textbf{(c)} After central He 
depletion but before C ignition.
\textbf{(d)} At central C depletion. \textbf{(e)} At central O depletion.
\textbf{(f)} At central Si depletion. \textbf{(g)} At onset of core 
collapse. \textbf{(h)} 100 s after core collapse (after explosive 
nucleosynthesis).}
\label{pfVsMass}
\end{figure*}

In Table~\ref{sonlySumm}, we compare our results for the
\textsl{s}-only elements, using the central values of the reaction
rates, with results of earlier calculations. The temperature dependence 
of the $^{79}$Se beta decay rate is not implemented in KEPLER; this may 
account for the lower value of the production factor for $^{80}$Kr compared to 
the results of \citealt{the07}. Our values of weak-s production percentage 
are significantly larger than those of \cite{rai93} and \cite{the07} 
presumably because of contributions 
from burning beyond O depletion. The present results have the correct trend, 
being smaller when the main \textsl{s}-process
is large.  Overall, however, the results for these central values of
the rates are surprisingly large.  But they are also very sensitive to
the reaction rates.  For example, for the L03 abundances, a
$15\,\%$ smaller value of the $^{12}$C($\alpha, \gamma$)$^{16}$O rate
gives weak-\textsl{s} contributions at least a factor of $2$ smaller
for these nuclei.  On the other hand, for the AG89 rates, the
weak-\textsl{s} contributions would be still larger.  This is yet
another example of the sensitivity of supernova nucleosynthesis to the
rates of the helium-burning reactions, and further evidence of the
need for more accurate rates.

\section{Evolution of Production Factors along the Stellar Burning History}

In this section, we compare the contribution of various stellar
burning phases to the \textsl{s}-process abundances for the
$25\,\Msun$ star, L03 initial abundances, and the central
values of the helium burning reaction rates: $R_{3\alpha}$ from
\cite{cau88}, and $R_{\alpha,12}$ equal to 1.2 times the rate
recommended by \cite{buc96}. Dump files are generated at specified
times during the life of the star as shown in Figure \ref{conv}.  From
these we extract the PFs for isotopes between $^{58}$Fe and $^{96}$Zr
and plot them in Figure~\ref{pfVsMass} versus their distance from the
stellar center in solar masses. The PF is defined as the mass fraction
of an isotope in a given stellar zone (or mass location) divided by
its solar mass fraction.  All the color bars of
Figures~\ref{pfVsMass}\emph{a}-\emph{h} have the same scale (from $1$
to $4 \times 10^{3}$) to facilitate comparison among them.
Figures~\ref{pfVsMass}\emph{a}-\emph{g} show the evolution of
production factors at various burning phases of the star, up to the
pre-supernova stage.  The black vertical line on the figures shows the
location of the initial mass cut (i.e., the position of the piston at
the base of the O burning shell right before the explosion).  Only the
stellar mass shells above that line have a chance of being ejected
during the explosion.  The gray shaded areas on figures
\ref{pfVsMass}\emph{f} and \ref{pfVsMass}\emph{g} show those regions
of the star where nucleosynthesis calculations are no longer being
performed because a nuclear statistical equilibrium network is
employed and these layers are known to become part of the iron core
and will not mix with the layers above before core collapse.  They
also lie inside the piston, i.e., below the pre-supernova mass cut.

\begin{figure*}
\centering
\includegraphics[angle=90,width=0.475\textwidth]{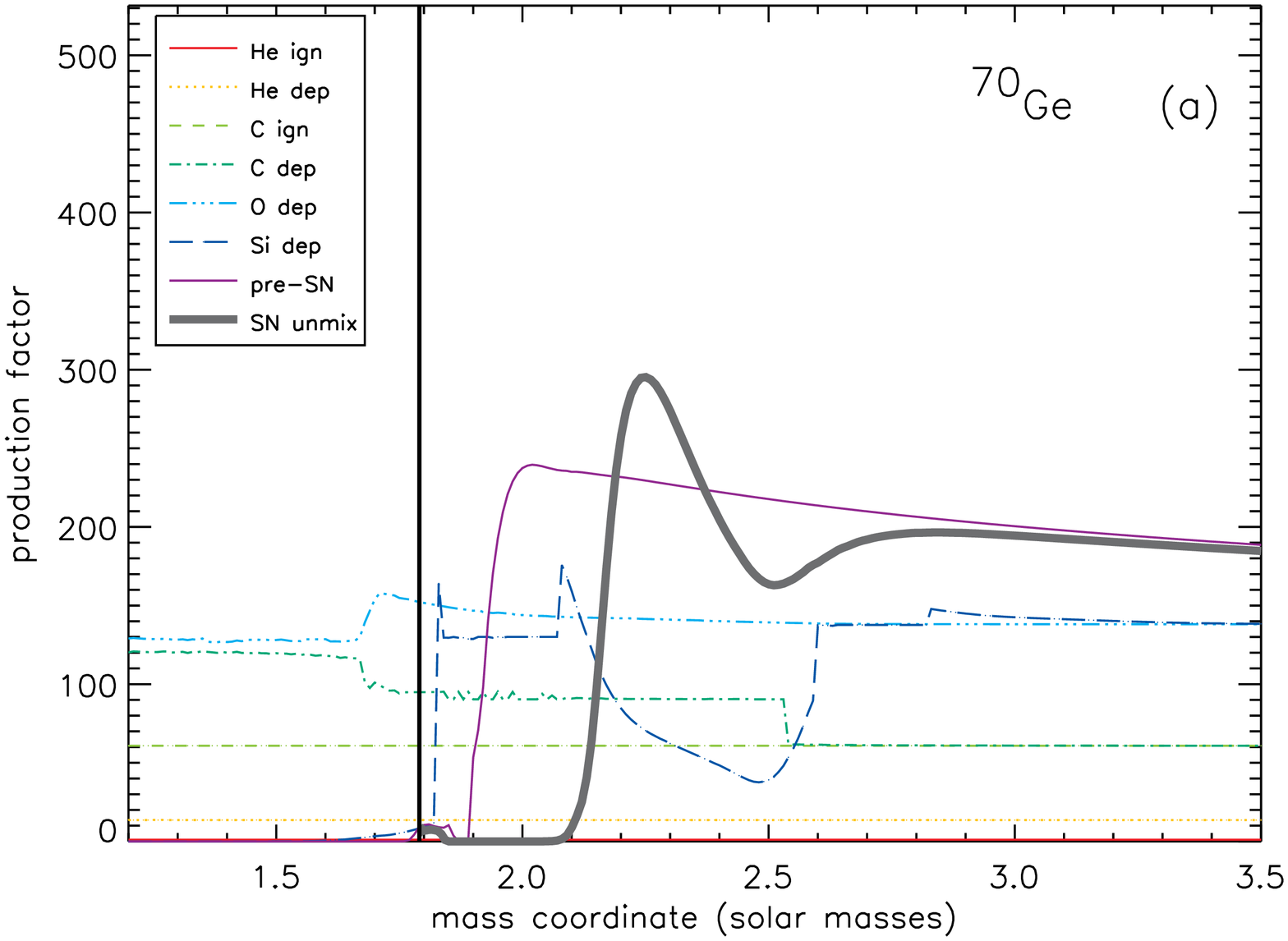}
\hfill
\includegraphics[angle=90,width=0.475\textwidth]{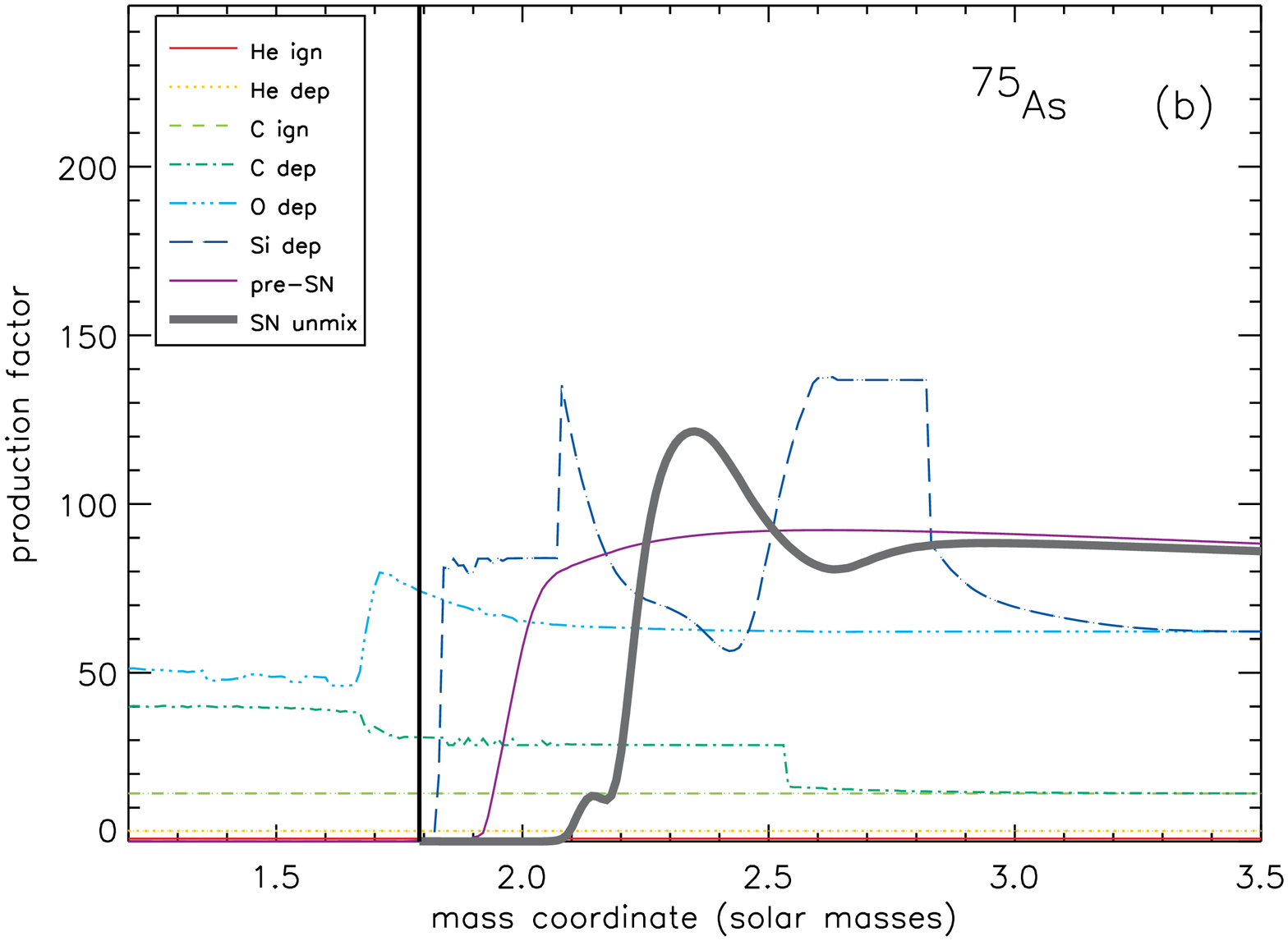}
\hfill
\includegraphics[angle=90,width=0.475\textwidth]{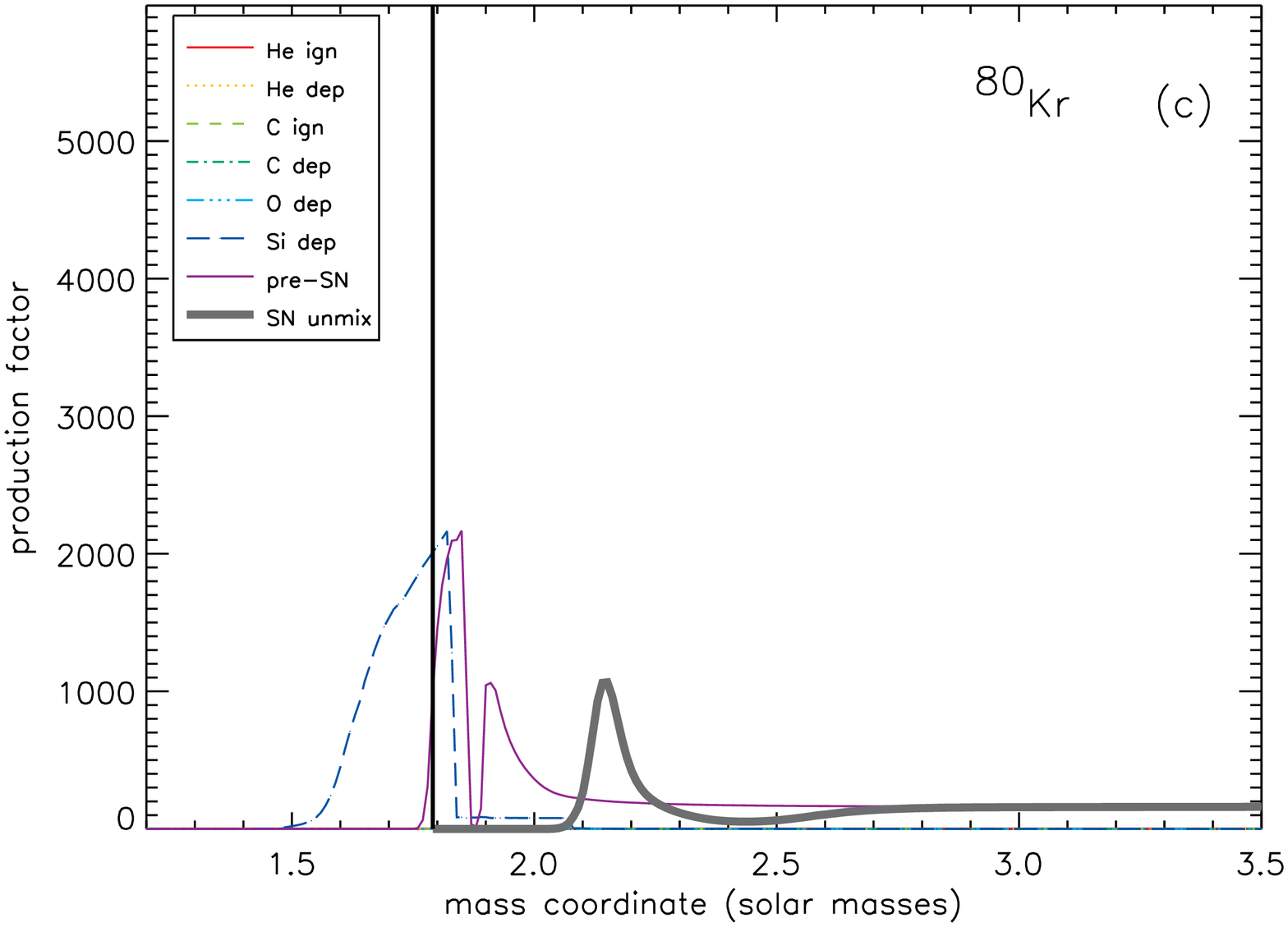}
\hfill
\includegraphics[angle=90,width=0.475\textwidth]{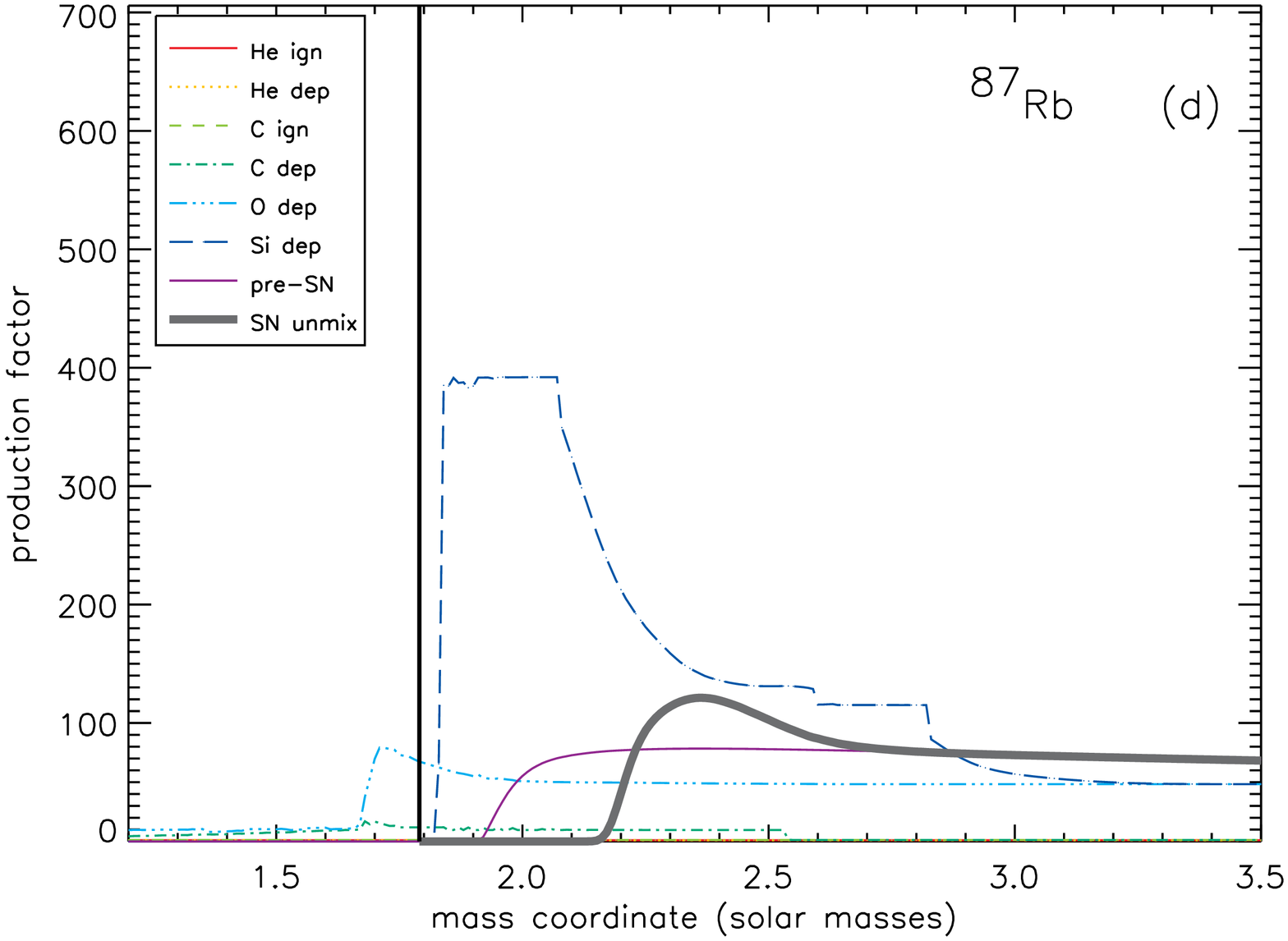}
\caption{Production factor versus the mass coordinate
  ($1.2\,\Msun-3.5\,\Msun$) for a $25\,\Msun$ star viewed at various
  times during the star's life for \textbf{(a)} $^{70}$Ge \textbf{(b)}
  $^{75}$As \textbf{(c)} $^{80}$Kr \textbf{(d)} $^{87}$Rb. The figures
  are plotted for the central values of the helium burning reaction
  rates. The thick vertical line shows the location of the initial
  mass cut.}
\label{singleIsotopes}
\end{figure*}

Below we give a detailed explanation for Figure~\ref{pfVsMass}:
\begin{itemize}
\item{Fig.~\ref{pfVsMass}\emph{a}:} at He ignition. Since no
  \textsl{s}-processing has yet occurred at this stage, the production 
factors are essentially zero.

\item{Fig.~\ref{pfVsMass}\emph{b}:} at He depletion (when the central
  He mass fraction has reached $1\,\%$). Central
  helium \textsl{s}-processing is in progress during this late He burning phase, as the temperatures become high
  enough to ignite $^{22}$Ne($\alpha,n$)$^{25}$Mg.

\item{Fig.~\ref{pfVsMass}\emph{c}:} just before central C ignition
  (when the central temperature reaches a value of $5\times10^8\,$K).
   Central helium burning has finished; this figure shows the final abundance and distribution of isotopes due to central helium
  burning.  About 1/3 of this \textsl{s}-process yield is located
  inside the later mass cut (compact remnant) of the supernova.  This
  has been largely neglected in most of the previous papers on the
  subject as it is hard to assess without simulation of the supernova.

\item{Fig.~\ref{pfVsMass}\emph{d}:} central C depletion (when the
  central temperature has reached $1.2\times10^9\,$K).  Central C
  burning contributes mostly to those regions of the star below the
  mass cut.  We also see the result of onset of the first carbon
 burning shell and of He shell burning (between mass coordinate of
 $6M_{\sun}$ and $7M_{\sun}$) which appears to be significant for most
 isotopes.

\item{Fig.~\ref{pfVsMass}\emph{e}:} at central O depletion (when the
  central O mass fraction has dropped below $5\,\%$).  Significant
  \textsl{s}-processing occurs during the C shell burning as seen in
  the enhancements to the PFs in the mass region above the mass
  cut.  There is also photodissociation during central O burning in the
  region below the mass cut.

\item{Fig.~\ref{pfVsMass}\emph{f}:} at Si depletion (when the central
  Si mass fraction drops below $10^{-4}$).  Additional
  photodissociation occurs during core Si burning.  Very strong
  overproduction occurs for some isotopes close to the mass cut.

\item{Fig.~\ref{pfVsMass}\emph{g}:} at the pre-supernova stage (when
  the contraction speed in the iron core reaches 1000
  km$\,$s$^{-1}$).

\item{Fig.~\ref{pfVsMass}\emph{h}:} $100\,$s after the passage of the
  shock wave.  This panel shows the PFs above the mass cut, when most
  nuclear reactions cease due to the very low temperatures.  The
  isotopes closest to the mass cut have essentially been destroyed
  with the exception of some local overproduction for isotopes below
  $^{70}$Ge.  These local enhancements can still lead to over-all
  increased supernova yields after mixing and fallback.
\end{itemize}

\begin{figure}
\centering
\includegraphics[angle=0,width=\columnwidth]{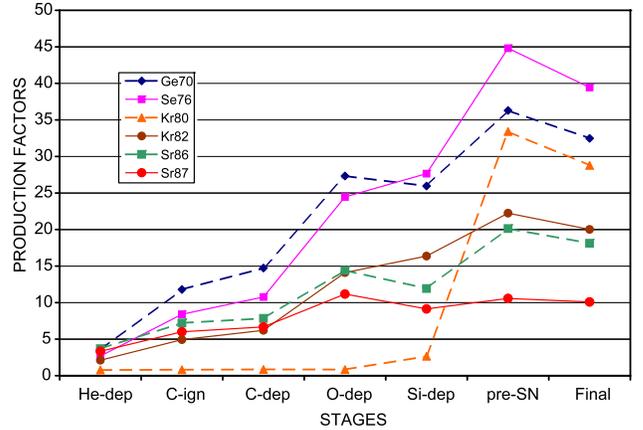}
\caption{Production factors of \textsl{s}-only nuclei lying outside
  the mass cut at various evolutionary stages for a $25\,\Msun$ star
  using the central values of the reaction rates and the L03
  abundances.}
\label{stages}
\end{figure}

In Figures~\ref{singleIsotopes}\emph{a}-\emph{d}, we show the
production factors of single isotopes ($^{70}$Ge, $^{75}$As,
$^{80}$Kr, and $^{87}$Rb) for a $25\,\Msun$ star versus the mass
coordinate ($1.2\,\Msun-3.5\,\Msun$) at various times during the
star's life, using the central values of the helium burning reaction
rates.  Figure~\ref{singleIsotopes}\emph{c} highlights the importance
of the late evolutionary stages: most of the $^{80}$Kr made during the
stages up to pre-supernova is destroyed and rebuilt during the
supernova explosion.  For $^{70}$Ge, $^{75}$As, and $^{87}$Rb, the C
shell burning and the pre-supernova stages both have significant
contributions for masses beyond about $2.5\,\Msun$. The temperature dependence of the $^{79}$Se beta decay
  rate is not implemented in KEPLER; the low values of the production
  factor for $^{80}$Kr in the earlier evolutionary stages presumably
  reflect this fact.  This comment also applies to Figure~\ref{stages}.  The lower values of the $^{22}$Ne($\alpha$, n)$^{25}$Mg rates used here may also contribute to smaller yields compared to some previous studies.

To assess the importance of various stages we have determined the PFs
for isotopes produced outside the mass cut at the times shown in
Figure~\ref{conv}.  It is clear from Figure~\ref{stages} that the
later evolutionary stages contribute significantly, especially for
$^{80}$Kr.  In the cases of $^{86,87}$Sr, however, little changes
after oxygen depletion.

\section{Sr, Y, Zr, and the LEPP Process}

As noted in the introduction, the LEPP process was introduced by
\cite{tra04} to explain the excess of the observed abundances of some
elements, especially Sr, Y, and Zr, above that produced by the
\textsl{r}-process, the main and weak \textsl{s}-processes, and the
\textsl{p}-process.  We examine here the uncertainties in the weak
\textsl{s}-process production of these elements.  The differences
shown in Table~\ref{minmax} and Fig.~\ref{pfVs12cag} for individual
isotopes appear much larger than the $8\,\% - 18\,\%$ differences that
led to the proposal of the LEPP process.  These differences, however,
are important only if, (1), the weak \textsl{s}-process makes a
sufficient fraction of an isotope, and if, (2), the differences
persist for the elemental differences, the observed quantities, not
just the differences in a single isotope.  We have calculated the
elemental production factors for the $25\,\Msun$ star and the
integrated result for the $15\,\Msun$, $20\,\Msun$ and $25\,\Msun$
stars, using the nominal reaction rates for $R_{3\alpha}$ and
$R_{\alpha,12}$ and the L03 abundances.  The results were
averaged over an IMF with a slope of $\gamma=-2.6$ (\citealt{sca86}) 
and divided by their solar mass fraction, as described in \cite{tur07}.  
We summarize the relevant quantities in Table~\ref{SrYZr}.  
For the $25\,\Msun$ star taken alone, the uncertainties are significant, 
$10\,\%-16\,\%$ compared to the $8\,\%-18\,\%$ differences that were to 
be explained by the LEPP process.  When one takes the averages over the
$15\,\Msun$, $20\,\Msun$, and $25\,\Msun$ stars, however,
uncertainties are smaller, $3\,\%-7\,\%$, but those for Sr are similar
to the $8\,\%$ LEPP effect.  All of these uncertainties are upper
limits, in the context of the present calculations, since they are the
extreme values for the entire set of reaction rates.  Differences in
PFs for these elements tend to cancel when one adds the production for
the three stars, but it is not known whether such cancellations will
occur for other elements.  And one must still consider the
uncertainties arising from uncertainties in the rates of the neutron
producing cross sections and the capture cross sections for various
poisons as described in the introduction.  Taken together, these
considerations leave the LEPP process only moderately robust.

\section{Conclusions}
We followed the entire nucleosynthesis throughout the life of massive
stars, from H-burning through Fe-core collapse, followed by a
supernova explosion.  We observe a strong sensitivity of the PFs for
weak-\textsl{s} process isotopes to $\pm 2\sigma$ variations of the
rates of the triple-$\alpha$ and of the
$^{12}$C($\alpha,\gamma$)$^{16}$O reactions.  This can be explained by
the variations in the PFs of medium-weight neutron poisons found by
\cite{tur07}, the changes in the amount of carbon left at the end of 
central helium burning, and the amount of $^{22}$Ne burnt
in central helium burning.

In most cases, our simulations yield lower PFs for the abundances of
L03 than for the abundances of AG89; the lower CNO
content of the L03 abundance set is responsible for the reduced
efficiency of the \textsl{s}-processing.  This tendency is not always
followed, however, as we see for the \textsl{s}-only nuclei with the
central rates, in Table~\ref{sonlySumm}.  The production of the
weak-\textsl{s} nuclei is highly sensitive to the rates of the helium
burning reactions.  In some cases (see the discussion of the results
of Table~\ref{sonlySumm}), we find that a $15\,\%$ change in these
rates may change the nucleosynthesis rates by more than a factor of
two.
 
We find that one must follow the entire evolution of the star to
evaluate accurately the contribution of massive stars to
\textsl{s}-process abundances.  Most earlier studies took into account
only the contribution of core He burning and shell C burning.  We show
that significant production takes place in later burning phases as
detailed for the \textsl{s}-only isotopes in Figure~\ref{stages}.  The
burning phases beyond shell C burning mostly destroy isotopes in the
core (especially near the mass cut), but overall lead to an increased
production of some isotopes.  The passage of the shock wave further
modifies those PFs, usually reducing them slightly.

We have examined the uncertainties in the weak-\textsl{s} process
production for Sr, Y, Zr owing to uncertainties in the helium-burning
reaction rates, and find that they are in the $3\,\%-7\,\%$ range.
This is smaller than the observed deficiencies of $8\,\%-18\,\%$ in
the known production mechanisms that led to the introduction of the
LEPP process.  On the other hand, we surely underestimate the total
uncertainties.  Uncertainties in the reaction rates of the neutron
producing and capture reactions are significant.  In addition,
uncertainties in the stellar models can have significant effects
(e.g., the treatment of hydrodynamics, convection, overshoot, etc.;
see \citealt{cos07} for a study of the effects of overshooting).  We
refer the reader to \cite{tur07} for a more complete discussion of
those approximations and physics uncertainties.  Combining all these
uncertainties may render the evidence for the LEPP process less
convincing.

We have shown that many aspects of nucleosynthesis in supernovae, the
production of the medium weight nuclei, as discussed in
\cite{tur07}, and that of the weak-\textsl{s}-process nuclei described
here, are highly sensitive to variations of the helium burning
reaction rates, within their experimental uncertainties.  This further
emphasizes the need for better values of the helium burning reaction
rates.

\acknowledgements

We thank Robert Hoffman for providing the solar abundance sets used in
this study and Stan Woosley for helpful discussions, including studies
on the relative influence of the two reaction rates. This research was
supported in part by the US National Science Foundation grants
PHY06-06007 and PHY02-16783, the latter funding the Joint Institute
for Nuclear Astrophysics (JINA), a National Science Foundation Physics
Frontier Center.  A.~Heger performed his contribution under the
auspices of the National Nuclear Security Administration of the US
Department of Energy at Los Alamos National Laboratory under contract
DE-AC52-06NA25396, and has been supported by the DOE Program for
Scientific Discovery through Advanced Computing (SciDAC;
DE-FC02-01ER41176).

\clearpage

\begin{deluxetable}{rrrrrr}
\tabletypesize{\scriptsize}
\tablecolumns{6} \tablecaption{Minimum and maximum values of the
  \textsl{s}-process production factors for a $25\,\Msun$ star with
  L03 initial abundances\label{minmax}}
\tablewidth{0pt}
\tablehead{
\colhead{}  &  \multicolumn{2}{c}{Within $2\sigma$ errors in $R_{\alpha,12}$}  &  \colhead{}   &  \multicolumn{2}{c}{Within $2\sigma$ errors in $R_{3\alpha}$}\\
\cline{2-3} \cline{5-6} \\
\colhead{Isotope}  & \colhead{Maximum} & \colhead{Minimum}  &
\colhead{}    &  \colhead{Maximum} & \colhead{Minimum}}
\startdata
$^{58}$Fe & 18.8 & 13.7 & & 19.1 & 13.1\\
$^{59}$Co & 15.6 & 9.3  & & 12.0 & 9.2 \\
$^{60}$Ni & 16.2 & 8.7  & & 11.9 & 6.0 \\
$^{61}$Ni & 36.1 & 20.7 & & 29.6 & 20.7\\
$^{62}$Ni & 52.5 & 22.8 & & 31.3 & 22.8\\
$^{64}$Ni & 43.8 & 21.0 & & 37.7 & 21.0\\
$^{63}$Cu & 35.0 & 14.1 & & 33.2 & 14.3\\
$^{65}$Cu & 34.5 & 18.5 & & 26.5 & 19.7\\
$^{64}$Zn & 4.4  & 2.3  & & 3.7  & 2.2 \\
$^{66}$Zn & 19.9 & 10.8 & & 20.0 & 10.5\\
$^{67}$Zn & 26.3 & 11.4 & & 22.9 & 11.4\\
$^{68}$Zn & 22.8 & 12.2 & & 20.0 & 12.2\\
$^{69}$Ga & 36.0 & 17.3 & & 32.6 & 21.8\\
$^{71}$Ga & 43.8 & 19.2 & & 40.7 & 19.2\\
$^{70}$Ge & 32.5 & 10.9 & & 36.3 & 13.8\\
$^{72}$Ge & 23.2 & 12.4 & & 28.7 & 11.9\\
$^{73}$Ge & 38.0 & 12.1 & & 30.2 & 11.7\\
$^{74}$Ge & 25.1 & 8.0  & & 16.8 & 8.0 \\
$^{75}$As & 30.4 & 13.4 & & 30.2 & 12.9\\
$^{76}$Se & 43.5 & 17.5 & & 54.3 & 13.9\\
$^{77}$Se & 35.8 & 12.2 & & 30.3 & 12.2\\
$^{78}$Se & 15.6 & 7.6  & & 20.0 & 6.4 \\
$^{80}$Se & 13.3 & 5.3  & & 10.4 & 5.3 \\
$^{79}$Br & 22.7 & 7.2  & & 17.8 & 9.3 \\
$^{81}$Br & 13.8 & 8.3  & & 16.7 & 6.7 \\
$^{80}$Kr & 42.6 & 4.7  & & 47.6 & 4.1 \\
$^{82}$Kr & 26.2 & 7.3  & & 29.9 & 5.5 \\
$^{83}$Kr & 12.7 & 5.3  & & 12.6 & 5.3 \\
$^{84}$Kr & 6.0  & 4.1  & & 6.6  & 3.8 \\
$^{86}$Kr & 48.4 & 3.1  & & 13.1 & 3.1 \\
$^{85}$Rb & 19.5 & 6.8  & & 16.9 & 6.8 \\
$^{87}$Rb & 20.5 & 7.4  & & 18.8 & 6.4 \\
$^{86}$Sr & 20.3 & 3.7  & & 23.9 & 2.9 \\
$^{87}$Sr & 15.3 & 2.3  & & 13.1 & 2.9 \\
$^{88}$Sr & 6.4  & 3.4  & & 5.8  & 3.3 \\
$^{89}$Y   & 4.7  & 2.6  & & 5.8  & 2.8 \\
$^{90}$Zr & 6.8  & 2.1  & & 8.0  & 1.9 \\
$^{91}$Zr & 3.1  & 1.6  & & 3.9  & 2.1 \\
$^{92}$Zr & 3.1  & 1.5  & & 3.2  & 1.5 \\
$^{94}$Zr & 2.4  & 1.2  & & 2.0  & 1.2 \\
$^{96}$Zr & 5.3  & 1.0  & & 3.3  & 1.0 \\
\enddata
\end{deluxetable}

\begin{turnpage}
\begin{deluxetable}{rccccccccc}
\tabletypesize{\scriptsize}
\tablecolumns{10}
\tablecaption{Percentage of \textsl{s}-only nuclei produced by the weak \textsl{s}-process for a $25\,\Msun$ star\label{sonlySumm}}
\tablewidth{0pt}
\tablehead{\colhead{} & \colhead{PF(X) AG} & \colhead{PF(X) The\tablenotemark{a}} & \colhead{PF(X) Lod} & \colhead{PF(AG)/PF(Lod)} & \colhead{PF(X)/PF($^{16}$O)AG} & \colhead{PF(X)/PF($^{16}$O)Lod} & \colhead{\%Weak-s AG\tablenotemark{b}} & \colhead{\%Weak-s Lod\tablenotemark{b}} & \colhead{\%Weak-s Raiteri\tablenotemark{c}}}
\startdata
$^{70}$Ge & 24.9 & 32.9	 & 32.5	& 0.8 & 2.49 & 2.12 & 124.7 & 106 & 64\\
$^{76}$Se & 38.7 & 29.2	 & 39.4	& 1.0 & 3.87 & 2.58 & 193.6 & 129 & 63\\
$^{80}$Kr & 18.5 & 47.7	 & 28.8	& 0.6 & 1.85 & 1.88 & 92.7  & 94  & 86\\
$^{82}$Kr & 20.9 & 28.4	 & 20.0	& 1.0 & 2.09 & 1.31 & 104.5 & 65  & 53\\
$^{86}$Sr & 7.5	 & 19.9	 & 18.1	& 0.4 & 0.75 & 1.18 & 37.4  & 59  & 24\\
$^{87}$Sr & 9.1	 & 15.8	 & 10.1	& 0.9 & 0.91 & 0.66 & 45.7  & 33  & 16\\
\enddata
\tablenotetext{a}{From \cite{the07}; results labelled 25K in Table 7.}
\tablenotetext{b}{Calculated  using Eq. 1.}
\tablenotetext{c}{From \cite{rai93}; Table 5.}
\end{deluxetable}
\end{turnpage}

\begin{deluxetable}{lrrrrr}
\tabletypesize{\scriptsize}
\tablecolumns{10}
\tablecaption{Fraction of Sr, Y, Zr produced in the weak-s process -- L03 abundances\label{SrYZr}}
\tablewidth{0pt}
\tablehead{\colhead{} & \colhead{PF-max} & \colhead{PF-min} & \colhead{f-max\tablenotemark{a}} & \colhead{f-min\tablenotemark{a}} & \colhead{$\Delta$f\tablenotemark{b}}}
\startdata
\\
\multicolumn{6}{c}{$25\,\Msun$ star:}\\
\\
Sr &	8.17 &	3.30 &	0.27 &	0.11 &	0.16\\
Y  &	5.76 &	2.64 &	0.19 &	0.09 &	0.10\\
Zr &	5.24 &	2.09 &	0.17 &	0.07 &	0.10\\

\\
\multicolumn{6}{c}{IMF average of $15\,\Msun$, $20\,\Msun$, and $25\,\Msun$ stars:}\\
\\
Sr &	4.402 &	2.201 &	0.144 &	0.072 &	0.072\\
Y  &	2.844 &	1.918 &	0.093 &	0.063 &	0.030\\
Zr &	3.089 &	1.519 &	0.101 &	0.050 &	0.051\\
\enddata
\tablenotetext{a}{Calculated from Eq.~\ref{PFeq}.}
\tablenotetext{b}{f-max - f-min: Spread of fractional production among calculated reaction rates.}
\end{deluxetable}

\end{document}